\documentclass[12pt,preprint]{aastex}
\usepackage{pslatex}
\usepackage{amssymb}

\shorttitle{Simulations of A Diffusive Shock with Multiple
Scattering Angular Distributions} \shortauthors{Wang X. and Yan Y.}
\def\jour#1#2#3#4{#1, {#2}, #3, #4}

\newcommand\lnp{{ Lecture Notes in Physics,\ }}

\def\jgr{ J. Geophys. Res.}

\def\ssr{ Space Sci. Rev.}
\def\apj{ Astrophys. J.}
\def\apjl{ Astrophys. J. Lett.}

\def\mnras{ M.N.R.A.S.}

\def\aap{ Astron. Astrophys.}

\newcommand\etal{{\it et al.~}}

\begin{document}

\title{The energy analysis for the monte carlo simulations of a diffusive shock}

\author{Xin Wang}

\affil{\textit{Key Laboratory of Solar Activities of National
Astronomical Observatories, Chinese Academy of Sciences,  Beijing
100012, China}}

\affil{\textit{State Key Laboratory of Space Weather, Chinese
Academy of Sciences, Beijing 100080, China}}

\author{Yihua Yan}

\affil{\textit{Key Laboratory of Solar Activities of National
Astronomical Observatories, Chinese Academy of Sciences,  Beijing
100012, China}}

\email{e-mail: wangxin@nao.cas.cn; yyh@nao.cas.cn}

\begin{abstract}
{According to the shock jump conditions, the total fluid's mass,
momentum, and energy should be conserved in the entire simulation
box. We perform the dynamical Monte Carlo simulations with the
multiple scattering law for energy analysis. The various energy
functions of time are obtained by monitoring the total particles'
mass, momentum, and energy in the simulation box. In conclusion, the
energy analysis indicates that the smaller energy losses in the
prescribed scattering law are, the harder the energy spectrum
produced is.}
\end{abstract}


\section{Introduction}
\label{Introduction}

The gradual solar energetic particles with a power-law energy
spectrum are generally thought to be accelerated by the first-Fermi
acceleration mechanism at the interplanetary shocks (IPs)
\citep{axford77,krymsky77,bell78,bo78}. It is well known that the
diffusive shock accelerated the particles efficiently by the
accelerated particles scattering off the instability of Alfven waves
which are generated by the accelerated particles themselves
\citep{gosling81,cvm90, Lee86,bkv03,plm06}. The diffusive shock
acceleration (DSA) is so efficient that the back-reaction of the
accelerated particles on the shock dynamics cannot be neglected. So
the theoretical challenge is how to efficiently model the full shock
dynamics \citep{bv06,ckvj10,zank00,li03,Lee05}. To efficiently model
the shock dynamics and the particles' acceleration, there are
largely three basic approaches: stationary Monte Carlo simulations,
fully numerical simulations, and semi-analytic solutions. In the
stationary Monte Carlo simulations, the full particle population
with a prescribed scattering law is calculated based on the
particle-in-cell (PIC) techniques \citep{ebj96,veb06}. In the fully
numerical simulations, a time-dependent diffusion-convection
equation for the CR transport is solved with coupled gas dynamics
conservation laws \citep{kj07,ZA10}. In the semi-analytic approach,
the stationary or quasi-stationary diffusion-convection equations
coupled to the gas dynamical equations are solved
\citep{bac07,mdv00}.  Since the velocity distribution of
suprathermal particles in the Maxwellian tail is not isotropic in
the shock frame, the diffusion-convection equation cannot directly
follow the injection from the non-diffusive thermal pool into the
diffusive CR population. So considering both the quasi-stationary
analytic models and the time-dependent numerical models, the
injection of particles into the acceleration mechanism is based on
an assumption of the transparency function for thermal leakage
\citep{bgv05,kj07,vainio07}.  Thus, the dynamical Monte Carlo
simulations based on the PIC techniques are expected to model the
shock dynamics time-dependently  and also can eliminate the
suspicion arising from the assumption of the injection
\citep{knerr96,wang11}. In plasma simulation (PIC and hybrid), there
is no distinction between thermal and non-thermal particles, hence
particle injection is intrinsically defined by the prescribed
scattering properties, and so it is not controlled with a free
parameter \citep{ckvj10}.

Actually, Wang \etal\cite{wang11} have  extended the dynamical Monte
Carlo models with an anisotropic scattering law. Unlike the previous
isotropic prescribed scattering law, the Gaussian scattering angular
distribution is used as the complete prescribed scattering law.
According to the extended prescribed scattering law, we obtained a
series of similar energy spectrums with little difference in terms
of the power-law tail. However, it is not clear how such a
prescribed scattering law can affect the particles' diffusion and
the shock dynamics evolution. To probe these problems, we expect to
diagnose the energy losses in the simulations by monitoring all of
the behaviors of the simulated particles.

In the time-dependent Monte Carlo models coupled with a Gaussian
angular scattering  law,  the results show that the total energy
spectral index and the compression ratio are both effected by the
prescribed scattering law. Specifically, the total energy spectral
index is an increasing function of the dispersion of the scattering
angular distribution, but the subshock's energy spectral index  is a
digressive function of the dispersion of the scattering angular
distribution \citep{wang11}. In the dynamical Monte Carlo
simulations, one find that it is the only way for the particles to
escape upstream via free escaped boundary (FEB). With the same size
of the FEB which limited the maximum energy of accelerated
particles, we find that different Gaussian scattering angular
distribution generate different  maximum energy particles through
the scattering process at the same simulation time.

In an effort to verify the efficiency of the energy transfer from
the thermal to superthermal and the effect of the shock dynamics
evolution on the shock structures, we perform a dynamical Monte
Carlo code on Matlab with Gaussian scattering angular distribution
by monitoring the particles' mass, momentum and energy as functions
of time. Our Gaussian scattering angular distribution algorithm
consists of four cases involving four specific standard deviation
values. This aim is to know if the various particle's loss functions
are dependent on the prescribed scattering law and the various kinds
of losses can directly determine the total compression ratio and
final energy spectral index with the same timescale of the shock
evolutions and the same size of FEB.

In Section \ref{sec-model}, the basic simulation method is
introduced with respect to the Gaussian scattering angular
distributions for monitoring the  particles' mass, momentum and
energy as function of time in each case. In Section
\ref{sec-results}, we present the shock simulation results and the
energy analysis for all cases with four assumptions of scattering
angle distributions. Section \ref{sec-summary} includes a summary
and the conclusions.

\section{Method}\label{sec-model}

The Monte Carlo model is a general model, although it is
considerably expensive computationally, and it is important in many
applications to include the dynamical effects of nonlinear DSA in
simulations. Since the prescribed scattering law in Monte Carlo
model instead of the field calculation in hybrid simulations
\citep{Giacalone04,wo96}, we assume that particle scatters
elastically according to a Gaussian distribution in the local plasma
frame and that the mean free path (mfp) is proportional to the
gyroradius (i.e., $\lambda \propto r_{g}$), where $r_{g}=pc/(qB)$,
and its value is proportional to its momentum. Under the prescribed
scattering law, the injection is correlated  with those ``thermal"
particles which manage to diffuse the shock front for obtaining
additional energy gains and become superthermal particles
\citep{ebg05}.

However, the basic theoretical limit to the accelerated particle's
energy arises from the accelerated particle's Larmor radius, which
must be smaller than the dimensions of the acceleration region by at
least $v_{s}/c$ \citep{hillas84}, where $v_{s}$ is the shock
velocity. The limitation to the maximum energy of accelerated
particles due to the large Larmor radius  would be ameliorated if
the scattering angular distribution is varied. In these simulations,
the size of the FEB is set as a finite length scale matched with the
maximum diffusive length scale. Here, we further investigate the
possibility that the accelerated particles scatter off the
background magnetic field in the acceleration region with not only
an isotropic scattering angular distribution, but also with an
anisotropic scattering angular distribution. Actually, this
anisotropic distribution would probably produce an important effect
on simulation results. With the same size of FEB, the scattering law
applied isotropic distribution or anisotropic distribution would
produce different maximum energy particles. So an anisotropic
scattering law in the theory of the CR-diffusion is also needed
\citep{bell04}.

\begin{figure}[h]\center
    \includegraphics[width=2.5in, angle=0]{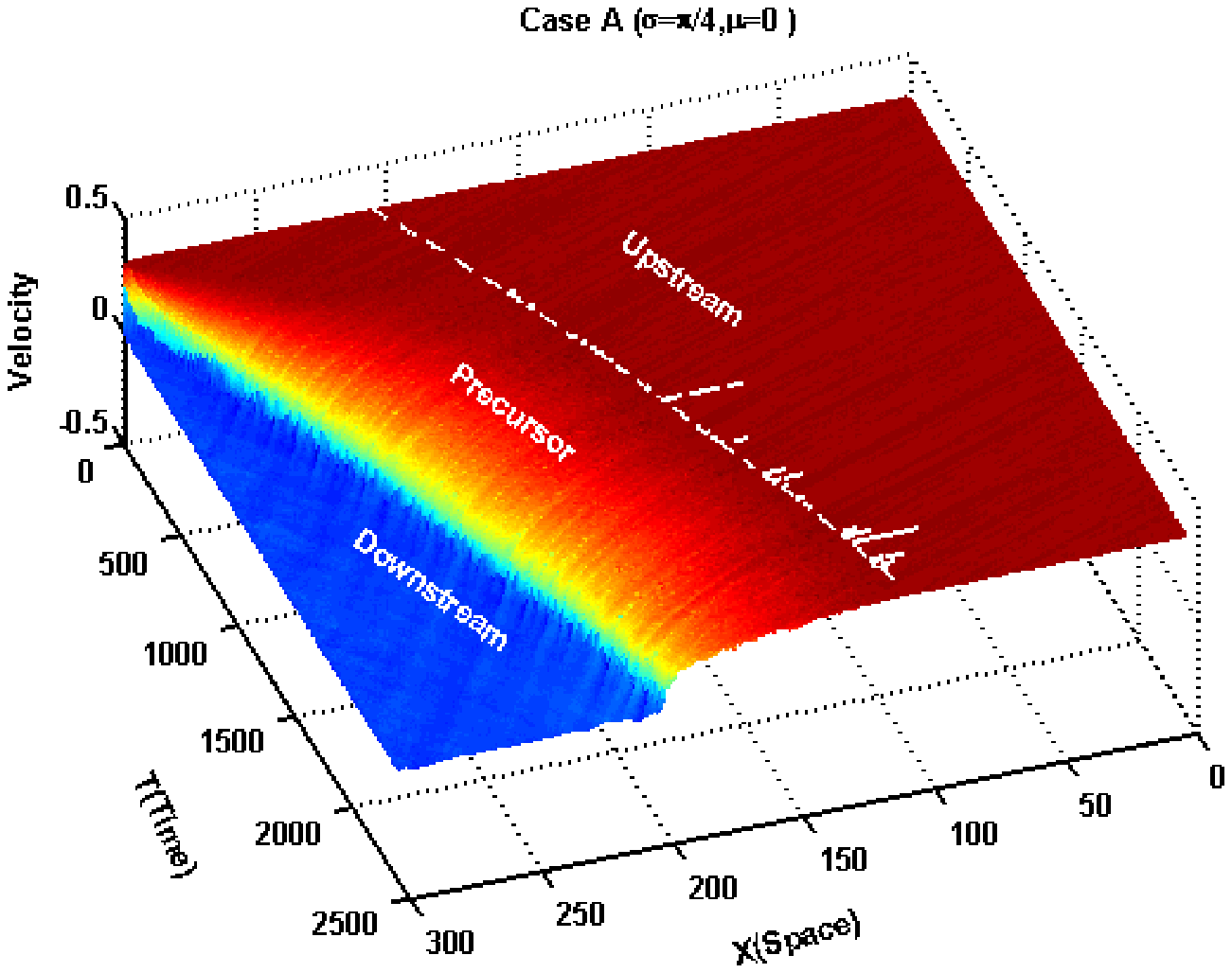}
    \includegraphics[width=2.5in, angle=0]{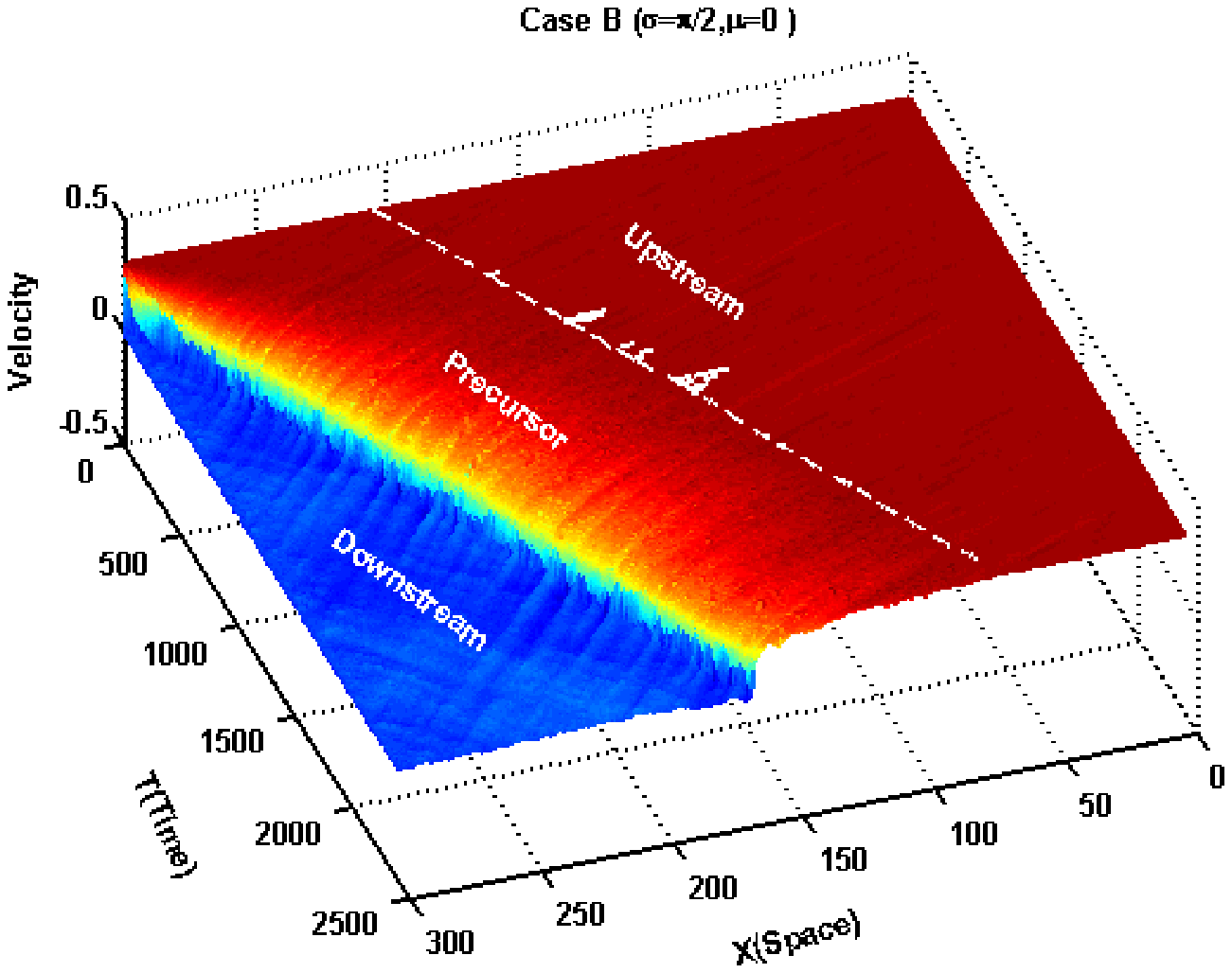}\\
    \includegraphics[width=2.5in, angle=0]{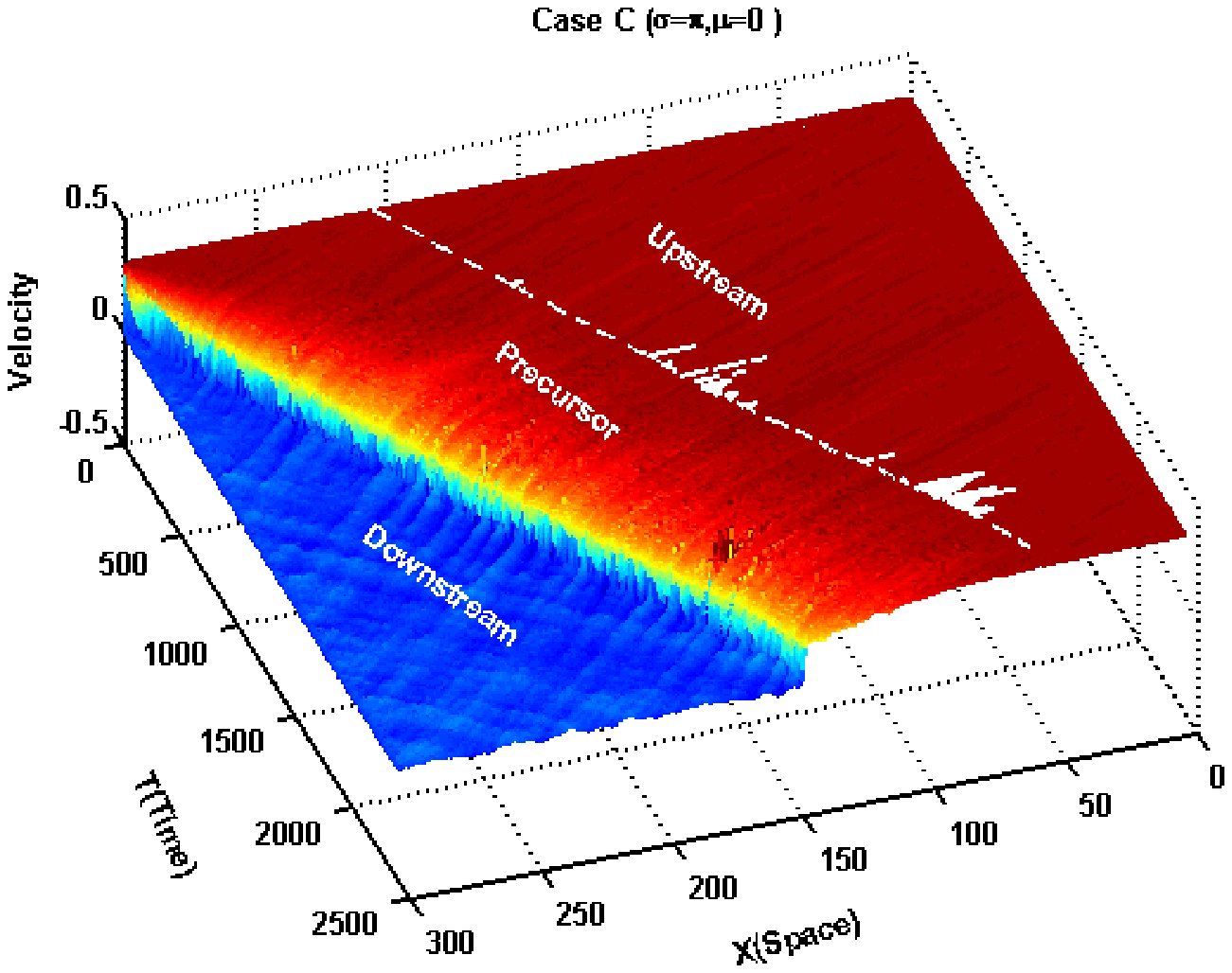}
    \includegraphics[width=2.5in,angle=0]{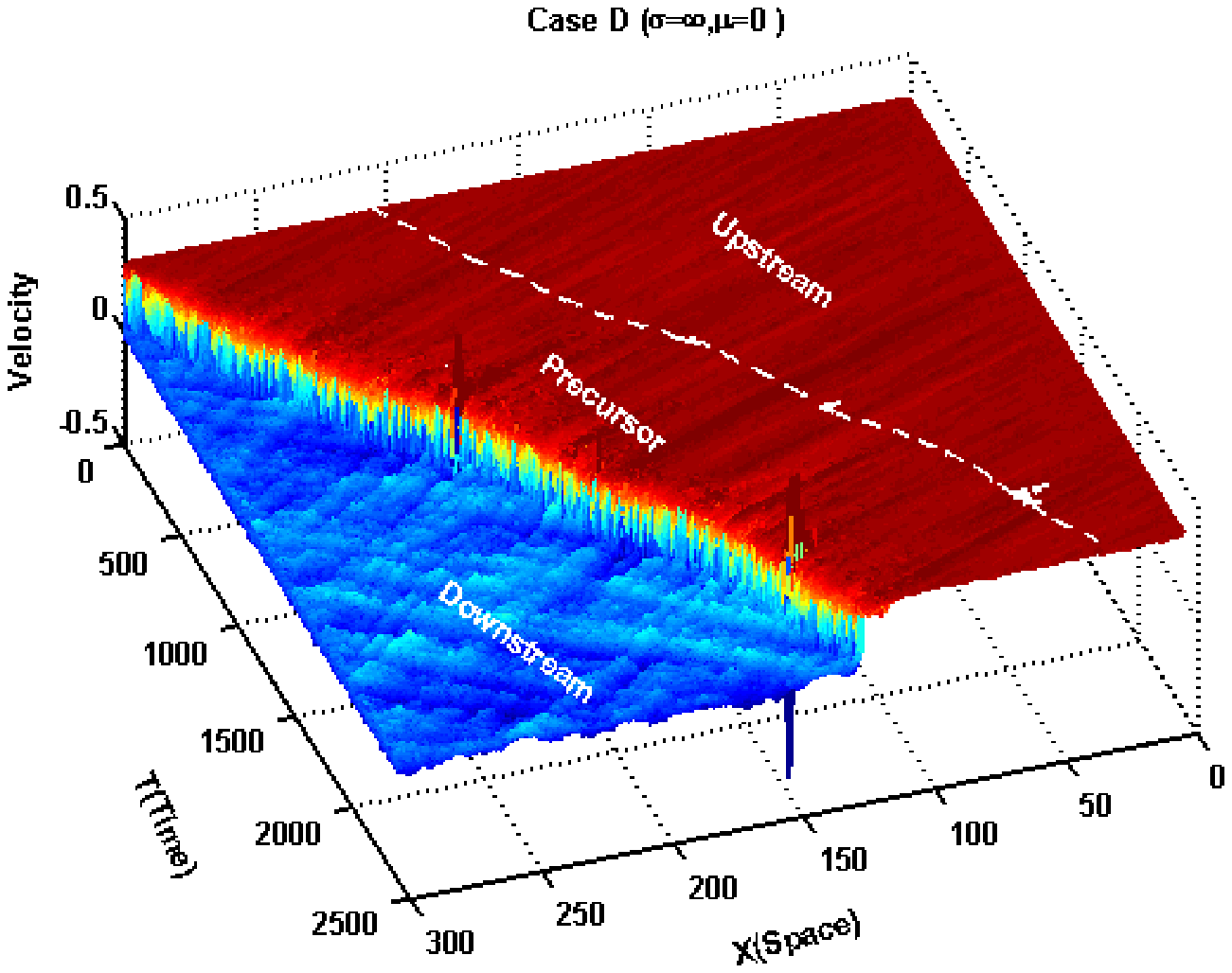}
\caption{The entire evolutional velocity profiles in four cases. The
dashed line denotes the FEB position in each plot. The precursor is
located in the area  between the downstream  region and the upstream
region in each case.}\label{fig:shock}
\end{figure}

The particle-in-cell techniques are applied in these dynamical Monte
Carlo simulations. The simulation box is divided up into some number
of cells and the field momentum is calculated at the center of each
cell \citep{forslund85,Spitkovsky08,nps08}. The total size of a
one-dimensional simulation box is set as $X_{max}$ and it is divided
into  the number of  grids $N_{max}$. Upstream bulk speed $U_{0}$
with an initial Maxwellian thermal velocity $V_{L}$ in their local
frame and the inflow in a ``pre-inflow box" (PIB) are both moving
along one-dimensional simulation box.  The parallel magnetic field
$B_{0}$ is along the $\hat{x}$ axis direction in the simulation box.
A free escaped boundary (FEB) with a finite size in front of the
shock position is used to decouple the escaped particles from the
system as long as the accelerated  particles beyond the position of
the FEB. The simulation box is a dynamical mixture of three regions:
upstream, precursor and downstream. The bulk fluid speed in upstream
is $U=U_{0}$, the bulk fluid speed in downstream is $U=0$, and the
bulk fluid speed with a gradient of velocity in the precursor region
is $U_{0}>U>0$. Because of the prescribed scattering law instead of
the particle's movement in the fields, the injected particles from
the thermalized downstream into the precursor for diffusive
processes are controlled by the free elastic scatter mechanism. To
obtain the information of the total particles in different regions
at any time, we build a database for recording the velocities,
positions, and time of the all particles, as well as the index and
the bulk speeds of the grids. The scattering angle distributions are
presented by Gaussian distribution function with a standard
deviation $\sigma$, and an average value $\mu$ involving four cases:
 (1) Case A: $\sigma=\pi$/4, $\mu=0$.
 (2) Case B: $\sigma=\pi$/2, $\mu=0$.
 (3) Case C: $\sigma=\pi$, $\mu=0$.
 (4) Case D: $\sigma=\infty$, $\mu=0$.
These presented simulations are all based on a one-dimensional
simulation box and the specific parameters are based on
\cite{wang11}.

\section{Results \& analysis}\label{sec-results}
We present the entire shock evolution with the velocity profiles of
the time sequences in each case as shown in Figure \ref{fig:shock}.
The total velocity profiles are divided into  three parts with
respect to the shock front and the FEB locations.  The upstream bulk
speed $U_{0}$ dynamically slows down by passing though the precursor
region, and  its value decreases to zero in the downstream region
(i.e. $U_{d}=0$). The precursor explicitly shows a different slope
of the bulk velocity and different final FEB locations in different
cases. The present velocity profiles are similar to the density
profiles in the previous simulations by \citet{wang11}, and the
different prescribed scattering law leads to the different shock
structures.

\begin{table*}[h]
\begin{center}
\caption{\label{tab:res}The Calculated Results}
\begin{tabular}{|c|c|c|c|c|c|c|c|c|c|c|c|c|c|c|c|c|c|c|c|c|c|}
  \hline   Items & Case A & Case B & Case C & Case D \\
%  \hline  $f(\delta\theta,\delta\phi)$ & Gaussian & Gaussian & Gaussian & Isotropy\\
 % \hline  $\sigma$ & $\pi/4 $& $\pi/2$ & $\pi$ & $\infty$ \\
 % \hline  $\mu $& 0 & 0 & 0 & 0 \\
  \hline  $M_{loss}$ & 1037 & 338 & 182 & 127 \\
  \hline  $P_{loss}$ & 0.0352 & 0.0189 & 0.0123 & 0.0084 \\
  \hline  $E_{loss}$ & 0.7468 & 0.5861 & 0.4397 & 0.3014 \\
  \hline  $E_{feb}$ & 0.8393 & 0.5881 & 0.5310 & 0.5022 \\
  \hline  $E_{in}$ & 1.5861 & 1.1742 & 0.9707 & 0.8036 \\
  \hline  $E_{tot}$ & 3.3534 & 3.4056 & 3.3574 & 3.4026 \\
  \hline  $R_{in}$ & 38.25\% & 25.67\% & 19.98\% & 14.80\% \\
  \hline  $R_{loss}$ & 22.27\% & 17.21\% & 13.10\% & 8.86\% \\
  \hline  $r_{tot} $& 8.1642 & 6.3532& 5.6753 & 5.0909 \\
  \hline  $r_{sub}$& 2.0975 & 3.0234 & 3.1998 & 3.9444 \\
  \hline  $\Gamma_{tot} $ & 0.7094 & 0.7802 & 0.8208 & 0.8667 \\
  \hline  $\Gamma_{sub} $ & 1.8668 & 1.2413 & 1.1819 & 1.0094 \\
  \hline  $v_{sh}$ & -0.0419 & -0.0560 & -0.0642 & -0.0733 \\
  \hline  $v_{sub}$ & 0.0805 & 0.1484 & 0.1613 & 0.2159 \\
  \hline  $V_{Lmax}$ & 11.4115 & 14.2978 & 17.2347 & 20.5286 \\
  \hline  $E_{peak}$ & 0.1650keV & 0.1723keV & 0.1986keV & 0.2870keV \\
  \hline  $E_{max}$ & 1.23MeV & 1.93MeV & 2.80MeV & 4.01MeV \\
  %\hline  $|V_{sh}|$ & 56.25km/s & 75.28km/s & 86.20km/s & 98.46km/s \\
  \hline
\end{tabular}
\end{center} %\tablefoot
{ The units of mass, momentum, and energy are normal to the proton
mass $m_{p}$, initial momentum $P_{0}$, and initial energy $E_{0}$,
respectively. The last two rows are shown as scaled values.}
\end{table*}
The various losses of the particles and the calculated results of
the shocks at the end of the simulation for the four cases are
listed in Table \ref{tab:res}. The initial box energy is $E_{0}$.
The subshock's compression ratio $r_{sub}$ and the total compression
ratio $r_{tot}$ are calculated from the fine velocity structures in
the shock frame in each case, respectively. The total energy
spectral index $\Gamma_{tot} $ and the subshock's energy spectral
index $\Gamma_{sub} $ are deduced from the corresponding total
compression ratio $r_{tot}$ and the subshock's compression ratio
$r_{sub}$ in each case. The $M_{loss}$, $P_{loss}$ and $E_{loss}$
are the mass, momentum, and energy losses which are produced by the
escaped particles via the FEB, respectively.

\begin{figure}[t]\center
    \includegraphics[width=2.5in, angle=0]{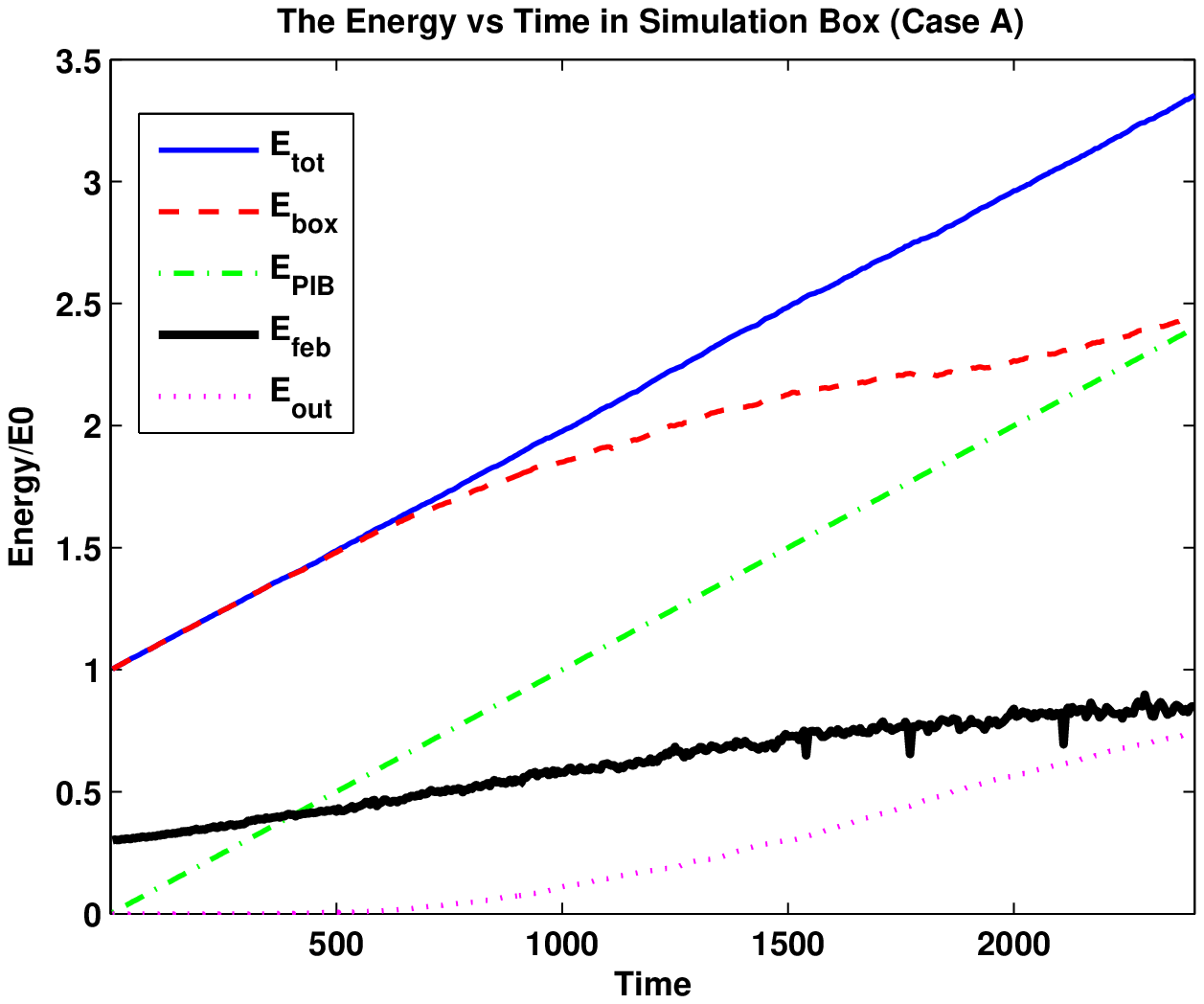}
    \includegraphics[width=2.5in, angle=0]{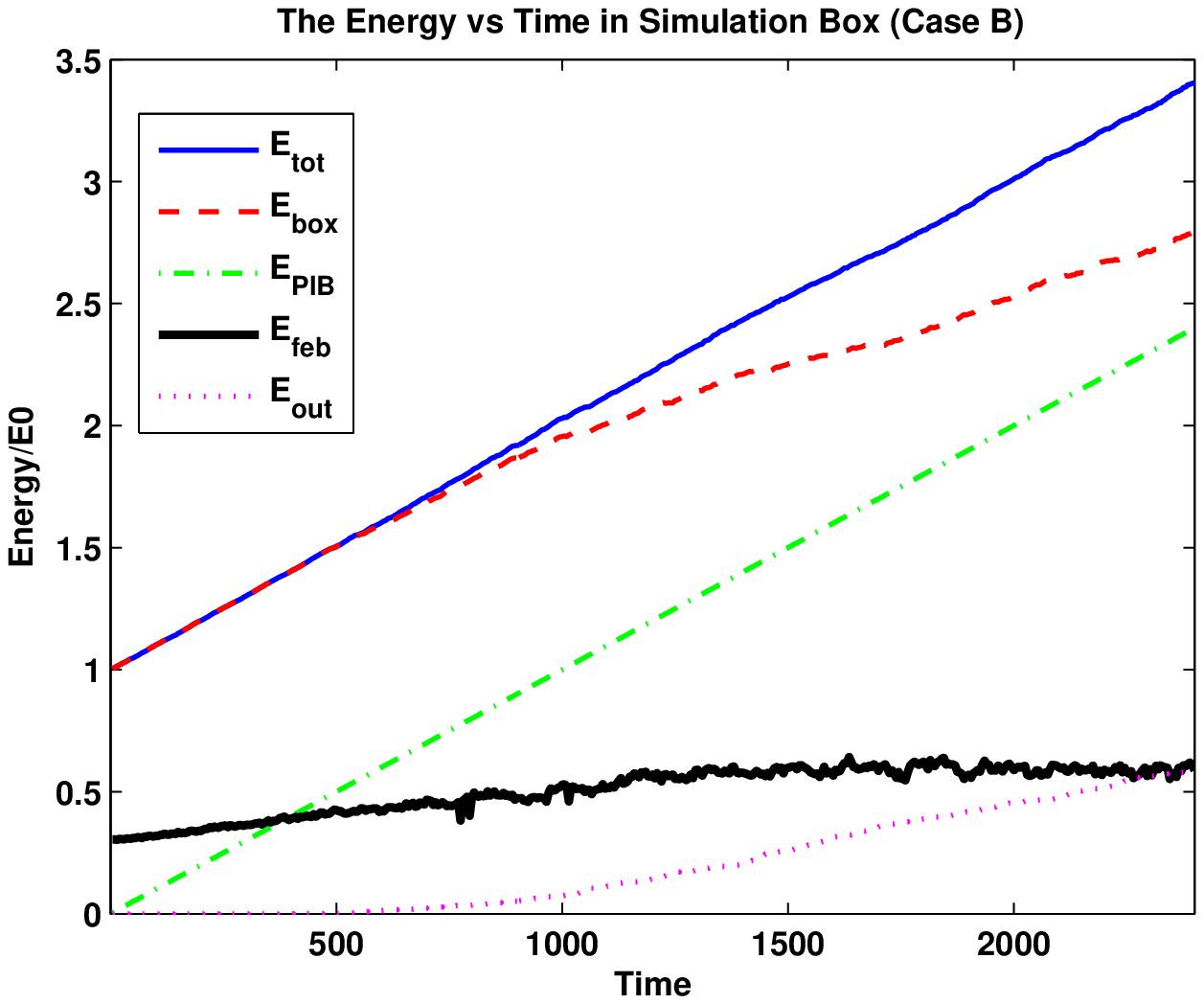}\\
    \includegraphics[width=2.5in, angle=0]{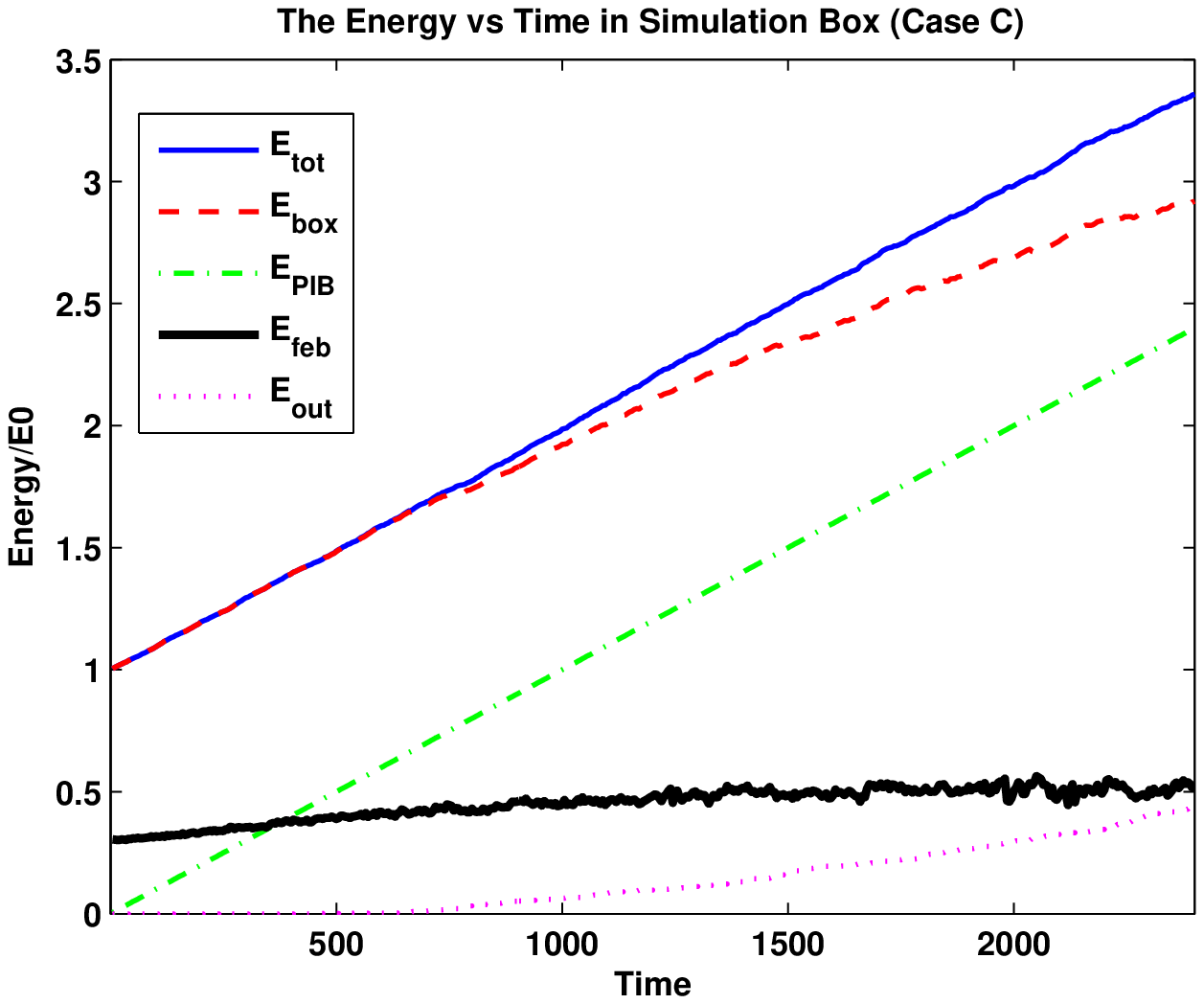}
    \includegraphics[width=2.5in,angle=0]{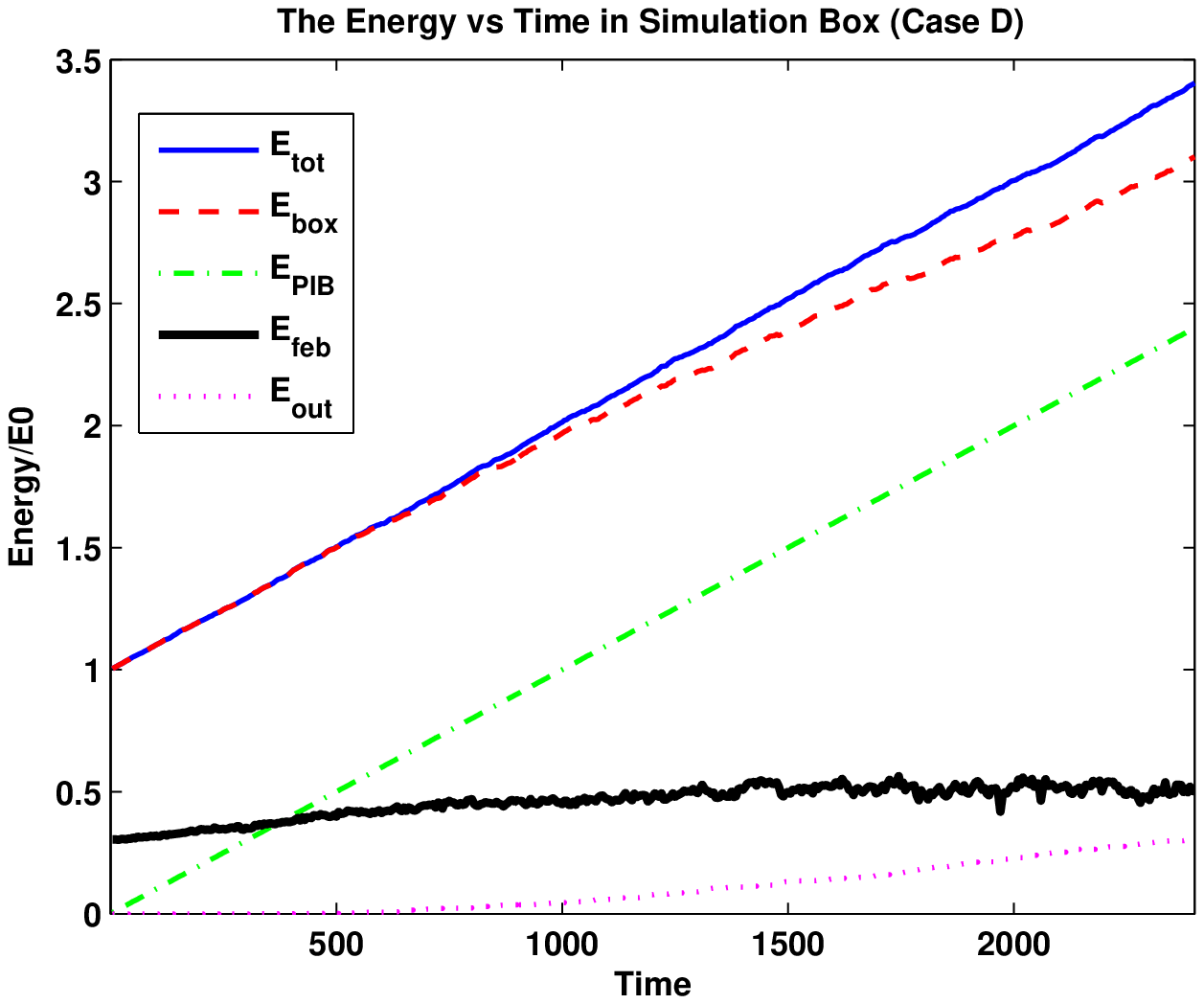}
  \caption{Various energy values vs. time (all normalized to the initial
   total energy $E_{0}$ in the simulation box) in each case. All quantities are calculated
    in the box frame.}\label{fig:eng}
\end{figure}

%\subsection{Energy Statistics}\label{subsec:es}
We have monitored  the mass, momentum and energy of the total
particles in each time step in each case. Figure \ref{fig:eng} shows
all the types of energy functions with respect to time. The total
energy $E_{tot}$ is the energy summation over the time in the entire
simulation box at any instant in time. The box energy $E_{box}$
presents the actual energy in the simulation box at any instant in
time. The supplement energy $E_{PIB}$ is the summation of the amount
of energy from the pre-inflow box (at the left boundary of the
simulation box) which enters into the simulation box with a constant
flux. The $E_{feb}$ presents a summation of the amount of energy
held in the precursor region. The $E_{out}$ indicates a summation of
the amount of energy which escapes via the FEB. Clearly, the total
energy in the simulation at any instant in time is not equal to the
actual energy in the box at any instant in time in each plot.

It can be shown that the incoming particles (upstream) decrease
their energy (as viewed in the box frame) as they scatter in the
shock precursor region. If each incoming particle loses a small
amount of energy as it first encounters the shock, this would
produce a constant linear divergence between the curves for
$E_{box}$ and $E_{tot}$. Actually, the cases in these simulations
produce  the non-linear divergence between the curves for $E_{box}$
and $E_{tot}$, consistent with Figure \ref{fig:eng}. Such behavior
is evident from individual particle's trajectories. Physically, all
the kinds of the losses occur in the precursor region owing to the
``back reaction" of the accelerated ions and the escaped particles
via FEB. Various energy functions obviously show that the case
applying the anisotropic scattering law produces a higher energy
loss, while the case applying the isotropic scattering law produces
a lower energy loss. Consequently, we consider that the prescribed
scattering law dominate the energy losses.

\subsection{Energy losses}
\begin{figure}[t]\center
   \includegraphics[width=2.5in, angle=0]{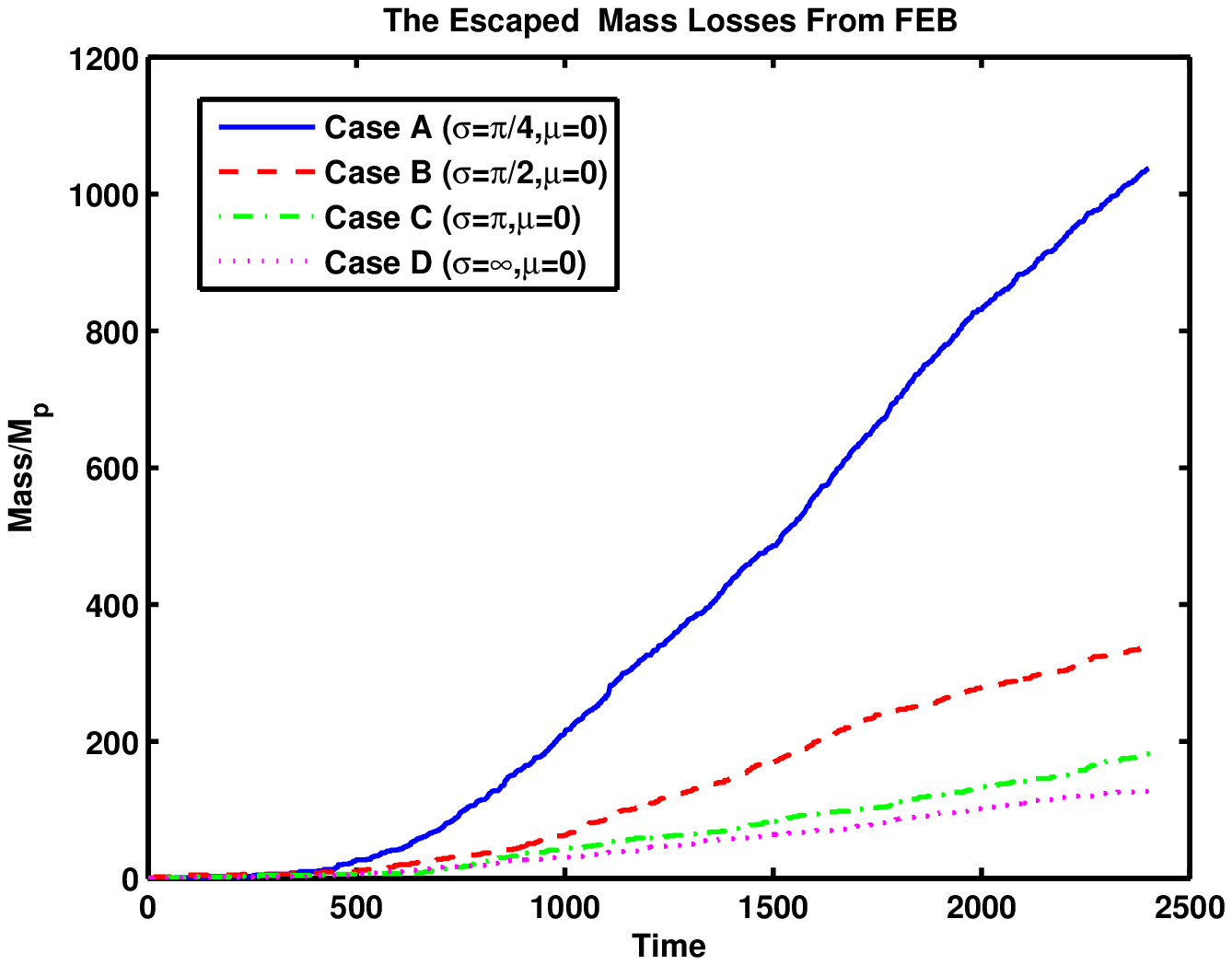}
    \includegraphics[width=2.5in, angle=0]{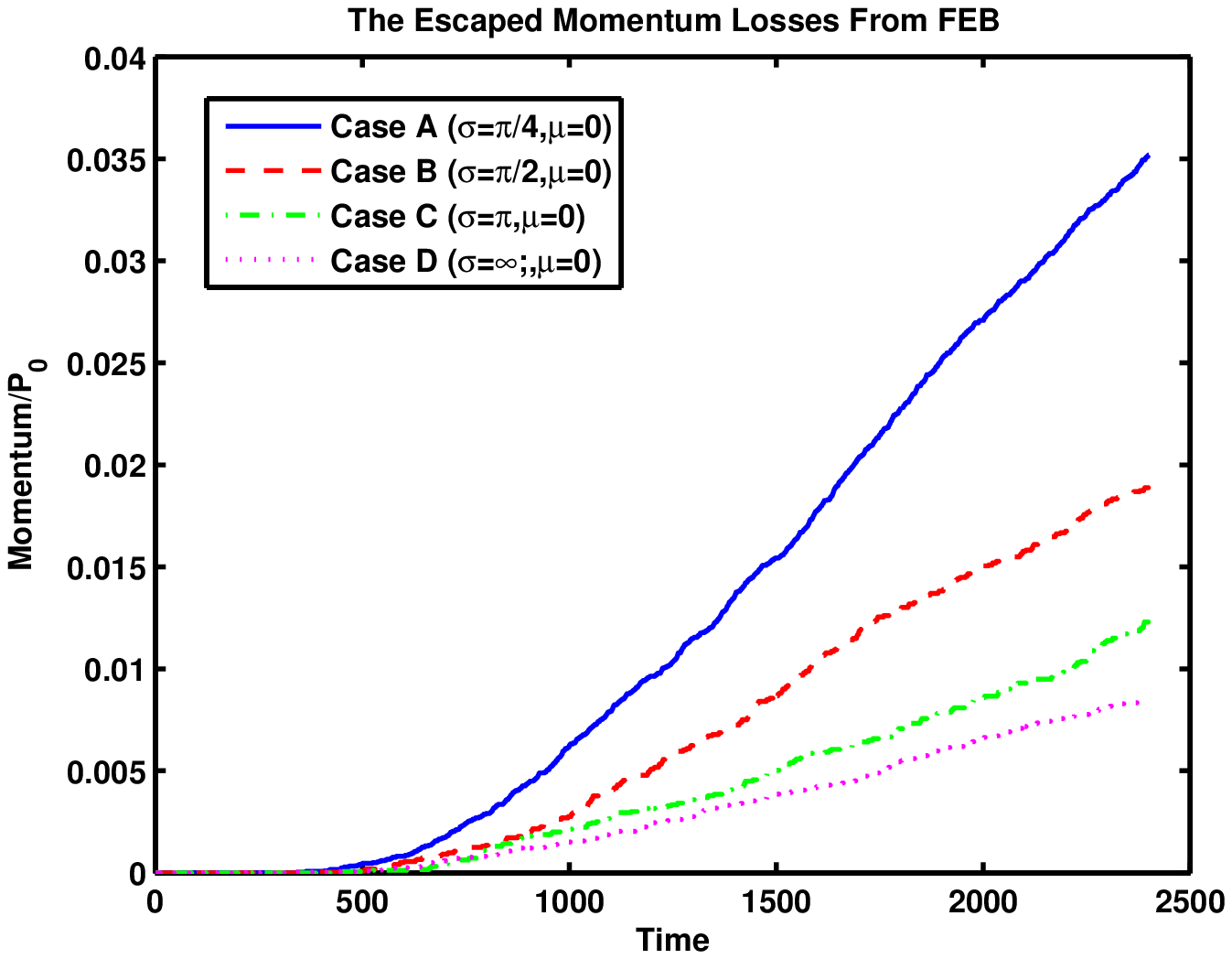}\\
   \includegraphics[width=2.5in, angle=0]{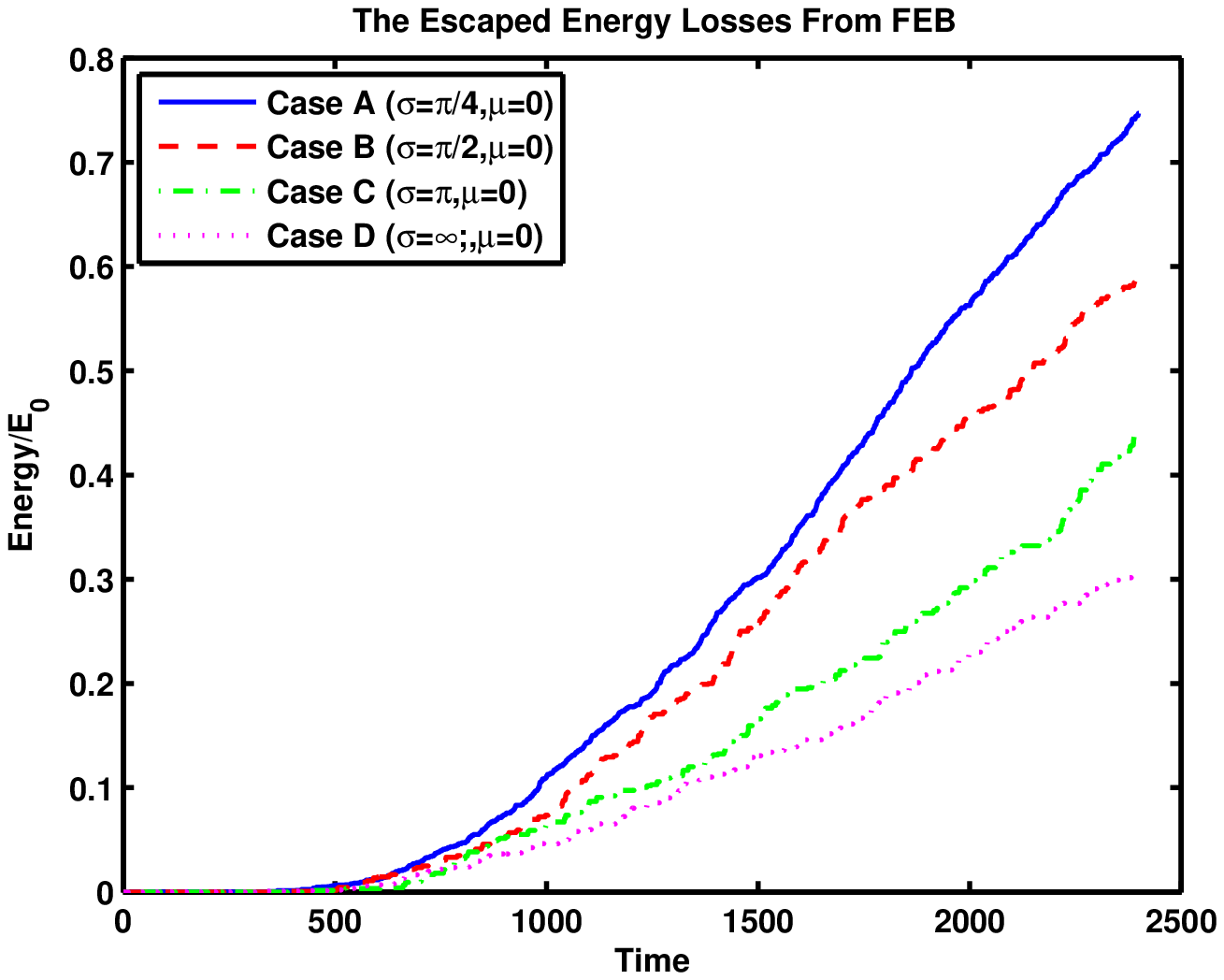}
   \includegraphics[width=2.5in, angle=0]{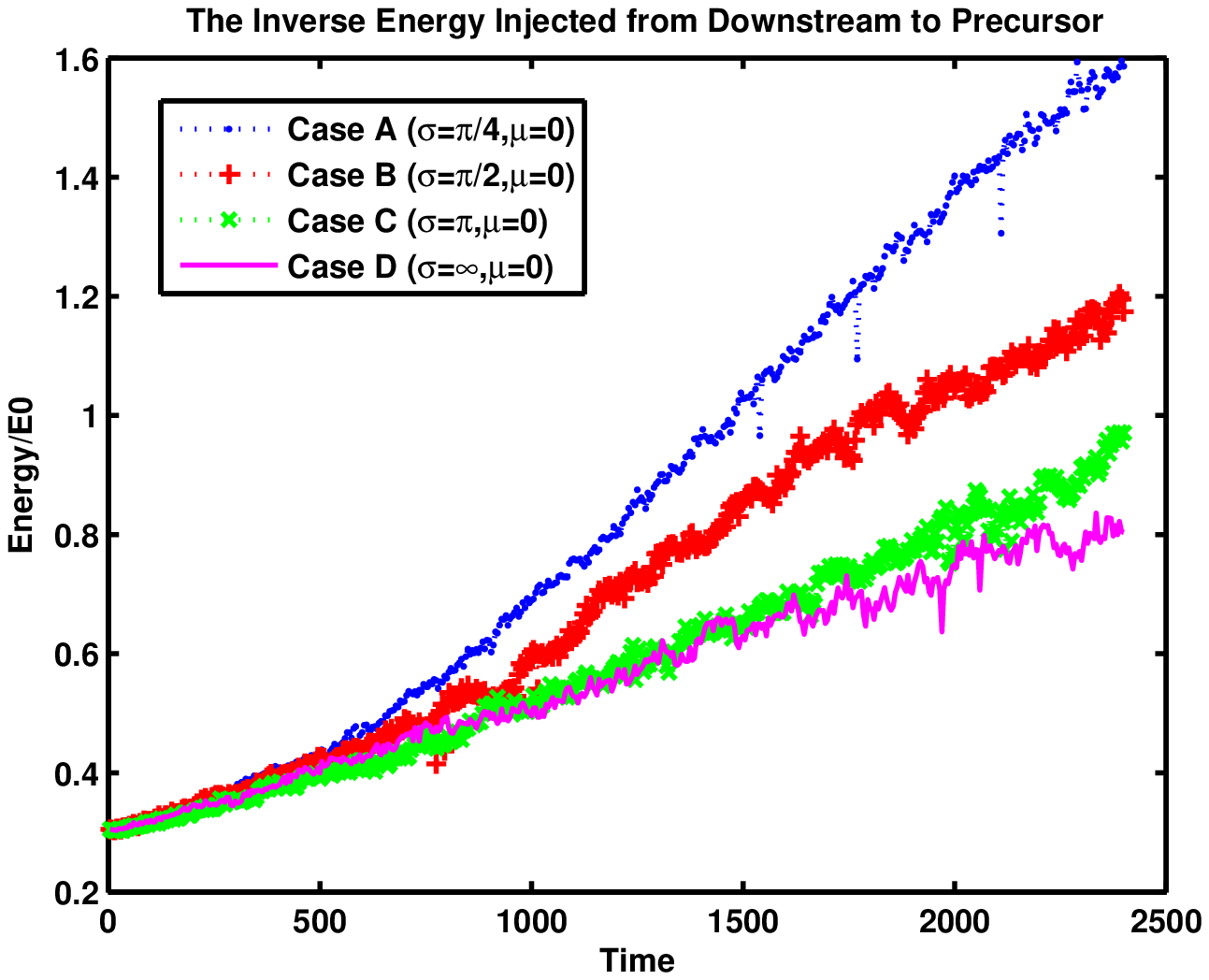}
\caption{ The four plots denote the mass losses, momentum losses,
energy losses and the inverse energy, respectively. The solid line,
dashed line, dash-dotted  line and the dotted line represent the
cases A, B, C and D in the first three plots. In the last plot, the
dashed lines marked with the signal of dot, plus, and cross, and the
solid line represent the cases A, B, C and D, respectively. The
units are normal to the proton mass $M_{p}$, initial total momentum
$P_{0}$ and initial total energy $E_{0}$,
respectively.}\label{fig:loss}
\end{figure}

With monitoring each particle in the grid of the simulation box in
any increment of the time, the escaped particles' mass, momentum,
and energy loss functions with the time are obtained and  shown in
Figure \ref{fig:loss}. Among these energy functions, the inverse
flow function is obtained from the injected particles from  the
thermalized downstream to the precursor.

Since the FEB is limited to be in front of the shock position with
the same size in each case,  once a single accelerated particle
moves backward to the shock beyond the position of the FEB, we
exclude this particle from the total system as the loss term in the
mass, momentum, and the energy conservation equations. According to
the Rankine-Hugoniot (RH) relationships,  the compression ratio of
the nonrelativistic shock with a large Mach number is not allowed to
be larger than the standard value of four \citep{Pelletier01}. Owing
to the energy losses which inevitably exist in the simulations, the
calculated results show a decreasing values of the loss of the mass,
momentum, and the energy from the cases A, B, and C to D,
respectively. Simultaneously, the inverse energy injected from
downstream to precursor is also a decreasing values of
$(E_{in})_{A}=1.5861$, $(E_{in})_{B}=1.1742$, $(E_{in})_{C}=0.9707$,
and $(E_{in})_{D}=0.8036$ from  the cases A, B, and C to D,
respectively.  The accurate energy losses are the values of
$(E_{loss})_{A}=0.7468$, $(E_{loss})_{B}=0.5861$,
$(E_{loss})_{C}=0.4397$, and $(E_{loss})_{D}=0.3041$ in each case.
The inverse energy is the summation of the energy loss $E_{loss}$
and the net energy $E_{feb}$ in the precursor (i.e.
$E_{in}=E_{loss}+E_{feb}$). So it is no wonder that the total shock
ratios are all larger than four because of the existence of energy
losses in all cases. Therefore, the difference of the energy losses
and the inverse energy can directly affect all aspects of the
simulation shocks.

\subsection{Subshock structure}\label{subsec:SS}
\begin{figure}[t]\center
    \includegraphics[width=2.5in, angle=0]{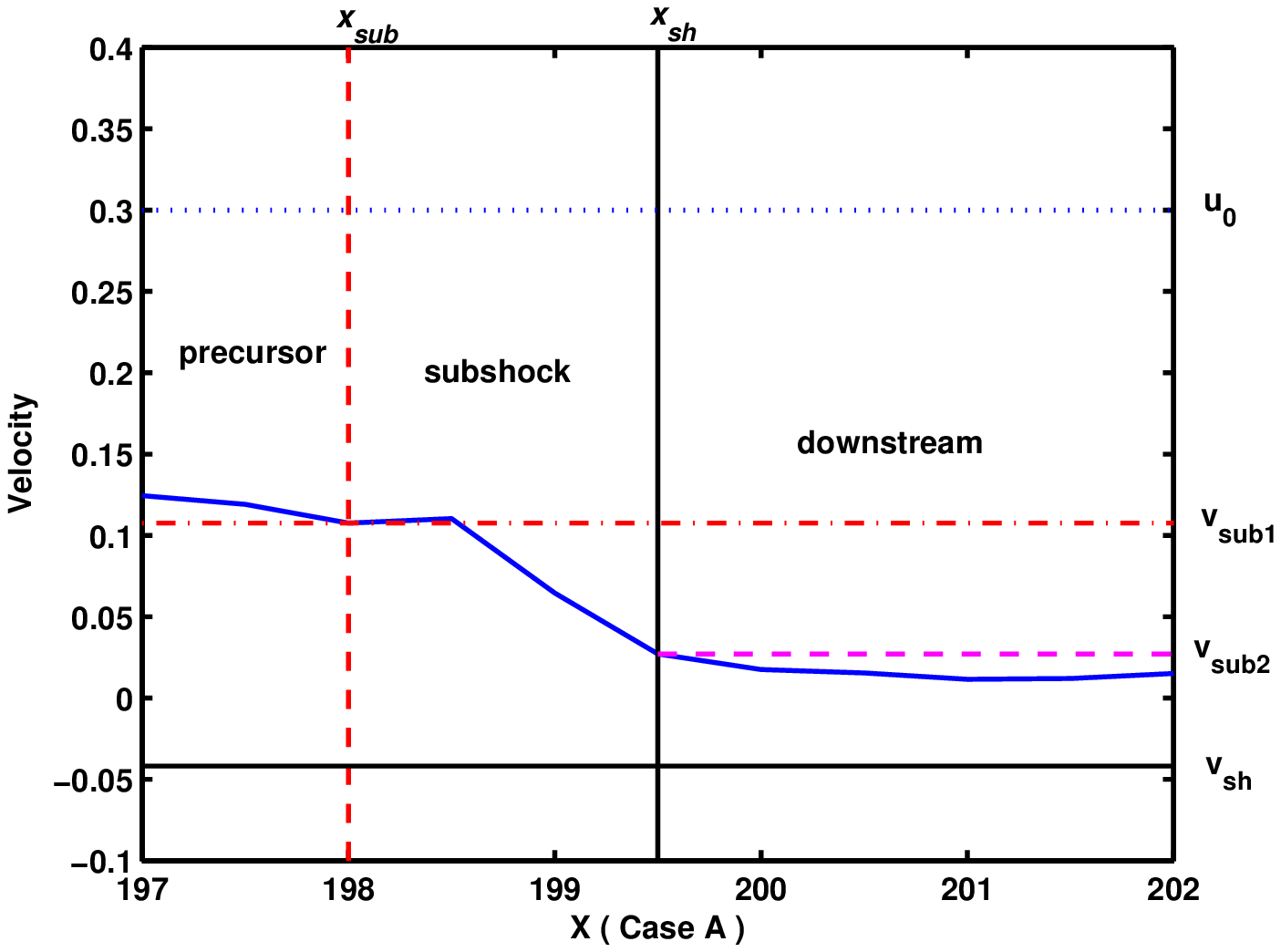}
    \includegraphics[width=2.5in, angle=0]{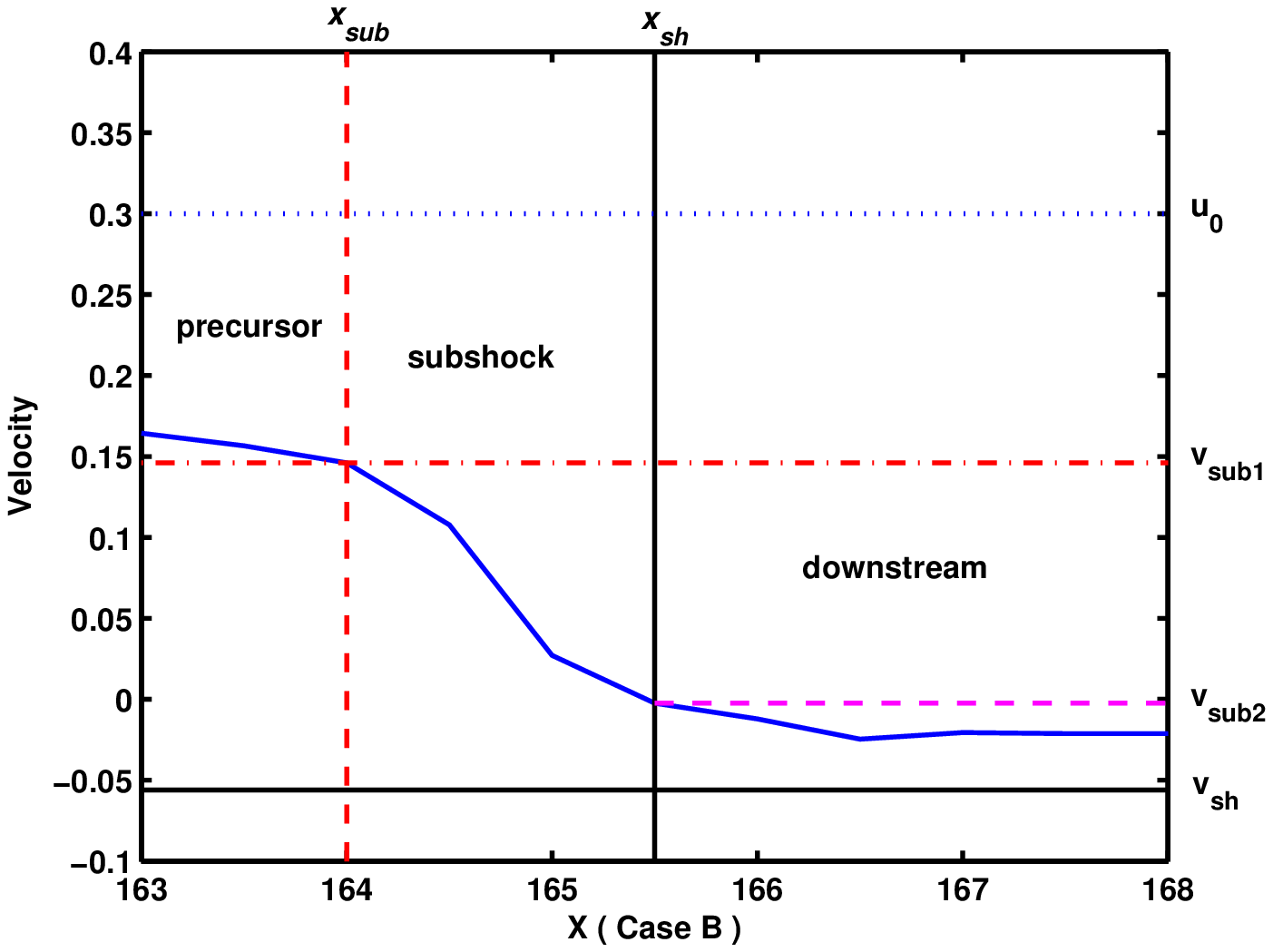}\\
    \includegraphics[width=2.5in, angle=0]{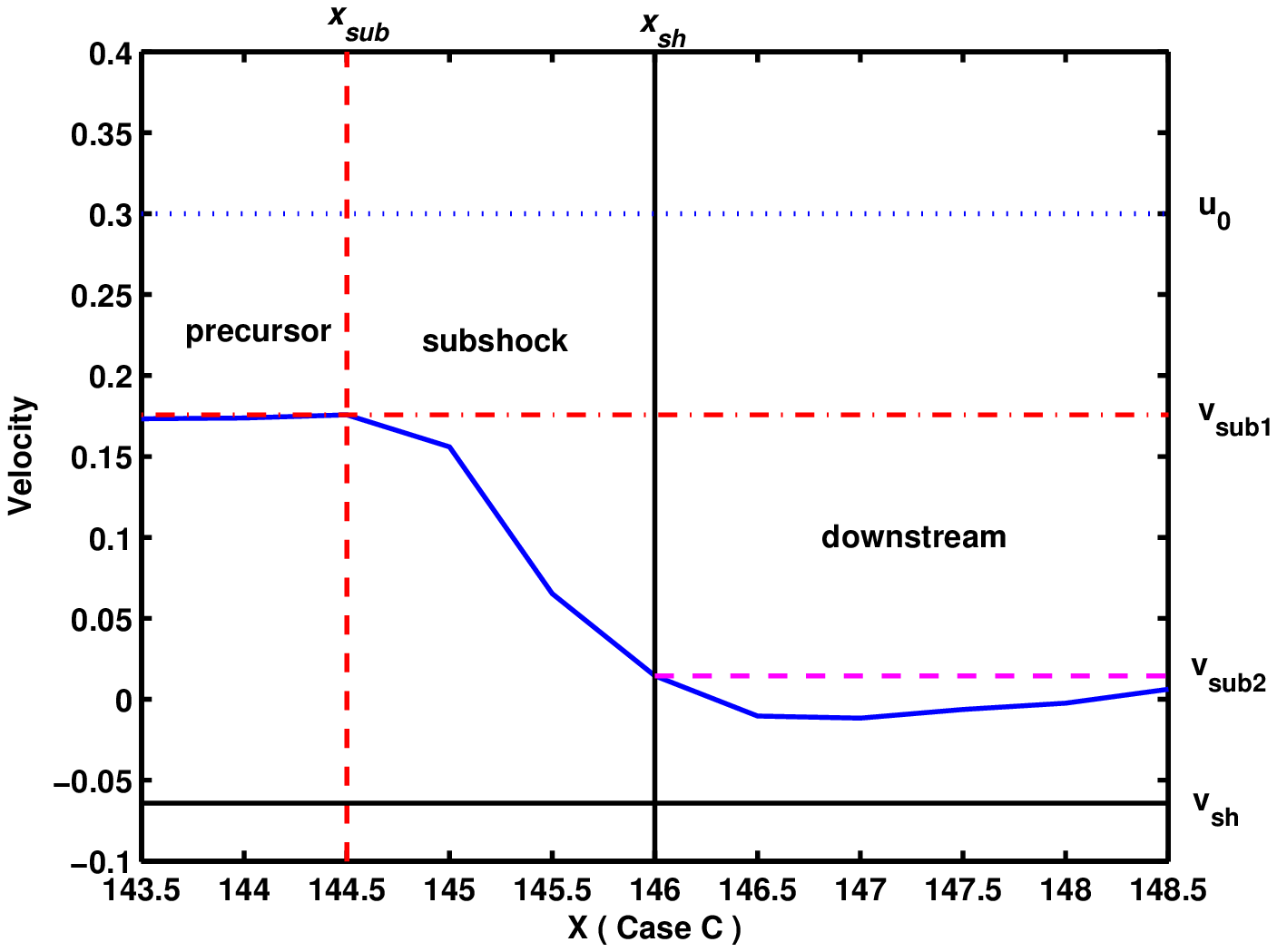}
    \includegraphics[width=2.5in, angle=0]{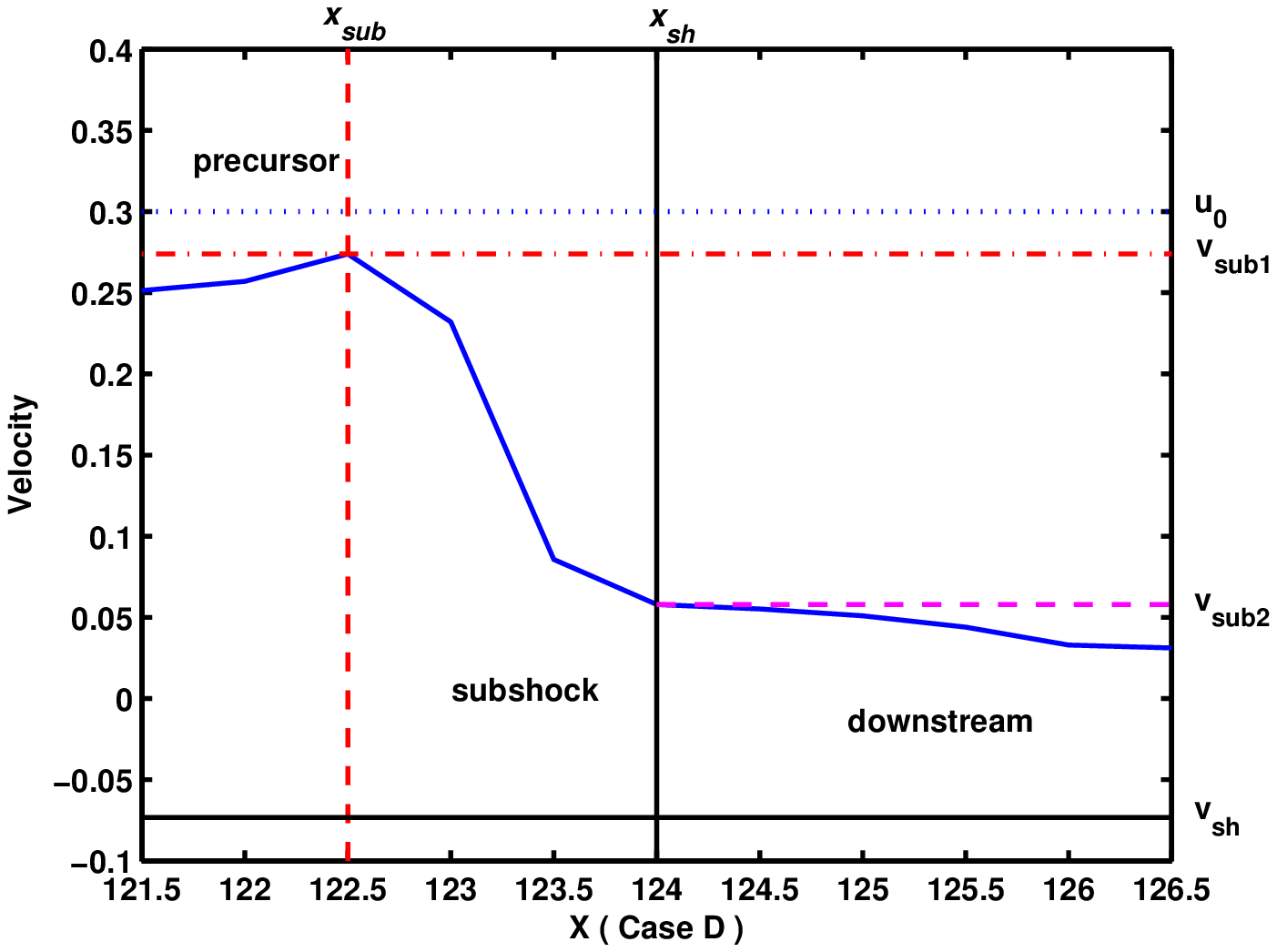}
\caption{Final subshock fine structures in the four cases. The
vertical solid and dashed lines indicate the positions of the shock
front and subshock in each plot, respectively. The horizontal solid,
dashed, dash-dotted and dotted lines show the values of the shock
velocity $v_{sh}$, subshock velocity $v_{sub2}$, subshock velocity
$v_{sub1}$ and initial bulk velocity $U_{0}$, respectively. Three
vertical blocks in each plot represent the three deflections of
velocity: precursor region, subshock region and downstream region.
All values of the velocity are based on the box frame.}
\label{fig:rsub}
\end{figure}

Figure \ref{fig:rsub} shows the subshock structure in each case at
the end of the simulation time. The specific structure in each plot
consists of three main parts: precursor, subshock and downstream.
The smooth precursor with a larger scale is between the FEB and the
subshock's position $X_{sub}$, where the bulk velocity gradually
decreases from the upstream bulk speed $U_{0}$ to $v_{sub1}$. And
the size of the precursor is almost equal to the diffusive length of
the maximum energy particle accelerated by the diffusion process.
The sharp subshock with a shorter scale only spans
three-grid-lengthsb involving a deep deflection of the bulk speed
abruptly decreasing from $v_{sub1}$ to $v_{sub2}$, where  the scale
of the three-grid-length is largely equal to the mean free path of
the averaged thermal particles in the thermalized downstream. The
value of the subshock's velocity can be defined by the difference
value of the two boundaries of the subshock.

\begin{equation}\label{eq:vsub}
v_{sub}= |v_{sub1}-v_{sub2}|.
\end{equation}

With the upstream bulk speed slowing down from  $U_{0}$ to zero, the
size of the downstream region is increasing with its constant shock
velocity $v_{sh}$ in each case, and the bulk speed is $U=0$ owing to
the dissipation processes which characterize the downstream. The gas
subshock is just an ordinary discontinuous classical shock embedded
in  the total shock with a comparably larger scale \citep{bere99}.

According to the subtle shock structures, the shock compression
ratio can be cataloged into two classes: one class presents the
entire shock named  the total compression ratio $r_{tot}$ and the
other class characterizes the subshock named the subshock's
compression ratio $r_{sub}$.  The values of the two kinds of
compression ratios can be reduced by the following formulas,
respectively.

\begin{equation}\label{eq:rtot}
r_{tot}=u_{1}/u_{2},\\
%r_{sub}= (v_{sub} +|v_{sh}|) /|v_{sh}|
\end{equation}
\begin{equation}\label{eq:rsub}
%r_{tot}=u_{1}/u_{2}\\
r_{sub}= (v_{sub} +|v_{sh}|) /|v_{sh}|,
\end{equation}
where $u_{1}=u_{0}+|v_{sh}|$ , $ u_{2}=|v_{sh}|$, and $u_{1}(u_{2})$
is the upstream (downstream) velocity in the shock frame, $v_{sub}$
is the subshock's velocity determined by Equation \ref{eq:vsub}, and
the shock velocity $v_{sh}$ at the end of the simulation is decided
by the following:
\begin{equation}
 v_{sh}=(X_{max}-X_{sh})/T_{max},\label{eq:vsh}
\end{equation} where $X_{max}$ is the total length of the simulation box,
$T_{max}$ is the total simulation time, and $X_{sh}$ is the position
of the shock at the end of the simulation. The specific calculated
results are shown in Table \ref{tab:res}. The values of the
subshock's compression ratios are $(r_{sub})_{A}$=2.0975 ,
$(r_{sub})_{B}$=3.0234, $(r_{sub})_{C}$=3.1998 and
$(r_{sub})_{D}$=3.9444 corresponding to the cases A, B, C and D,
respectively. The total shock compression ratios with the values of
$(r_{tot})_{A}$=8.1642, $(r_{tot})_{B}$=6.3532,
$(r_{tot})_{C}$=5.6753, and $(r_{tot})_{D}$=5.0909  also correspond
to the cases A, B, C and D, respectively. In comparison, the total
shock compression ratios are all larger than the standard value four
and the subshock's compression ratios are all lower than the
standard value four. Additionally, the value of  the total shock
compression ratio decreases from  Cases A, B, and C to D, while the
subshock's compression ratio increases from  Cases A, B, and C to D.
These differences are naturally attributed to the different fine
subshock structures.

\subsection{Maximum energy}
\begin{figure}[t]\center
    \includegraphics[width=2.5in, angle=0]{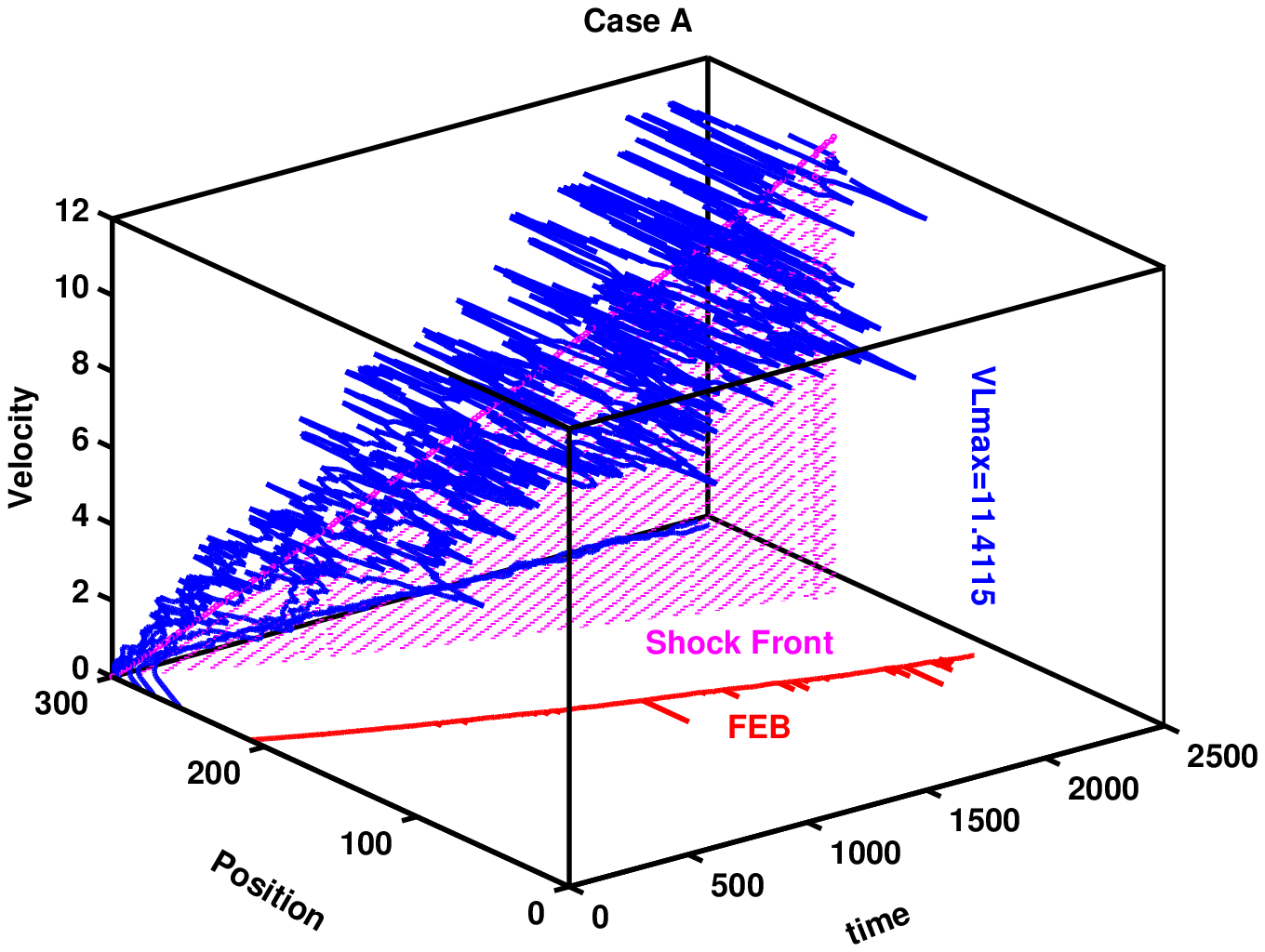}
    \includegraphics[width=2.5in, angle=0]{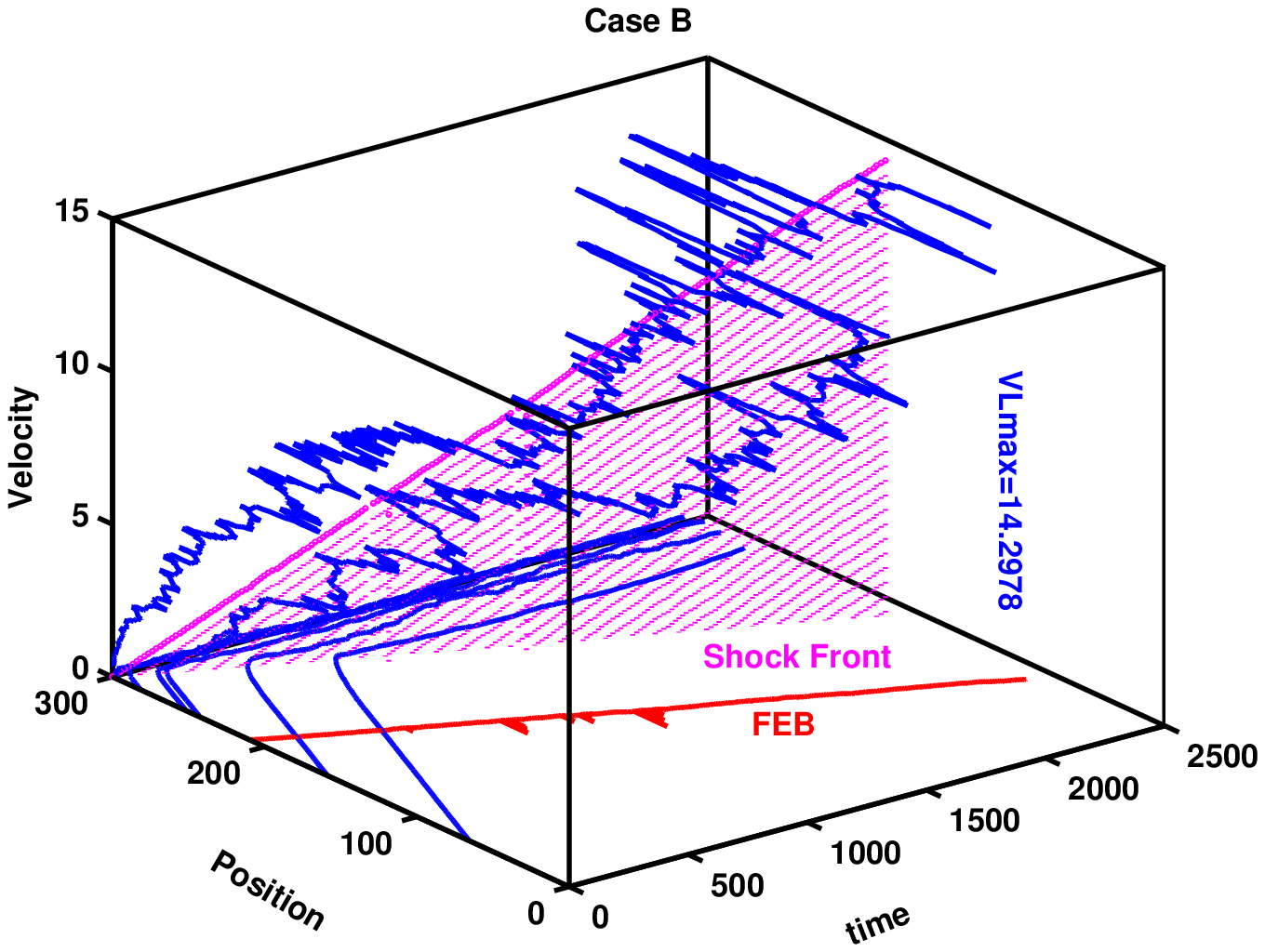}\\
    \includegraphics[width=2.5in, angle=0]{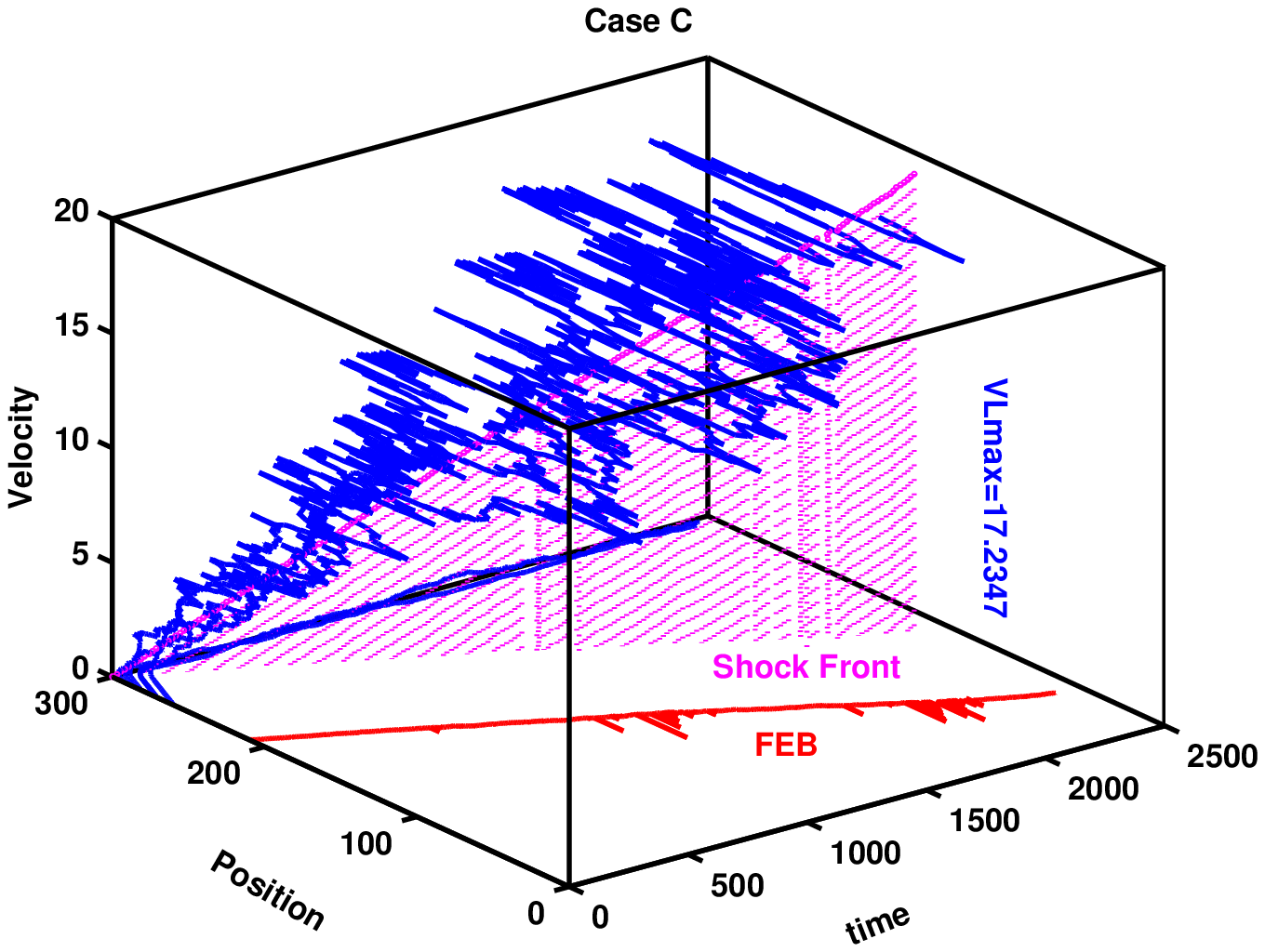}
    \includegraphics[width=2.5in, angle=0]{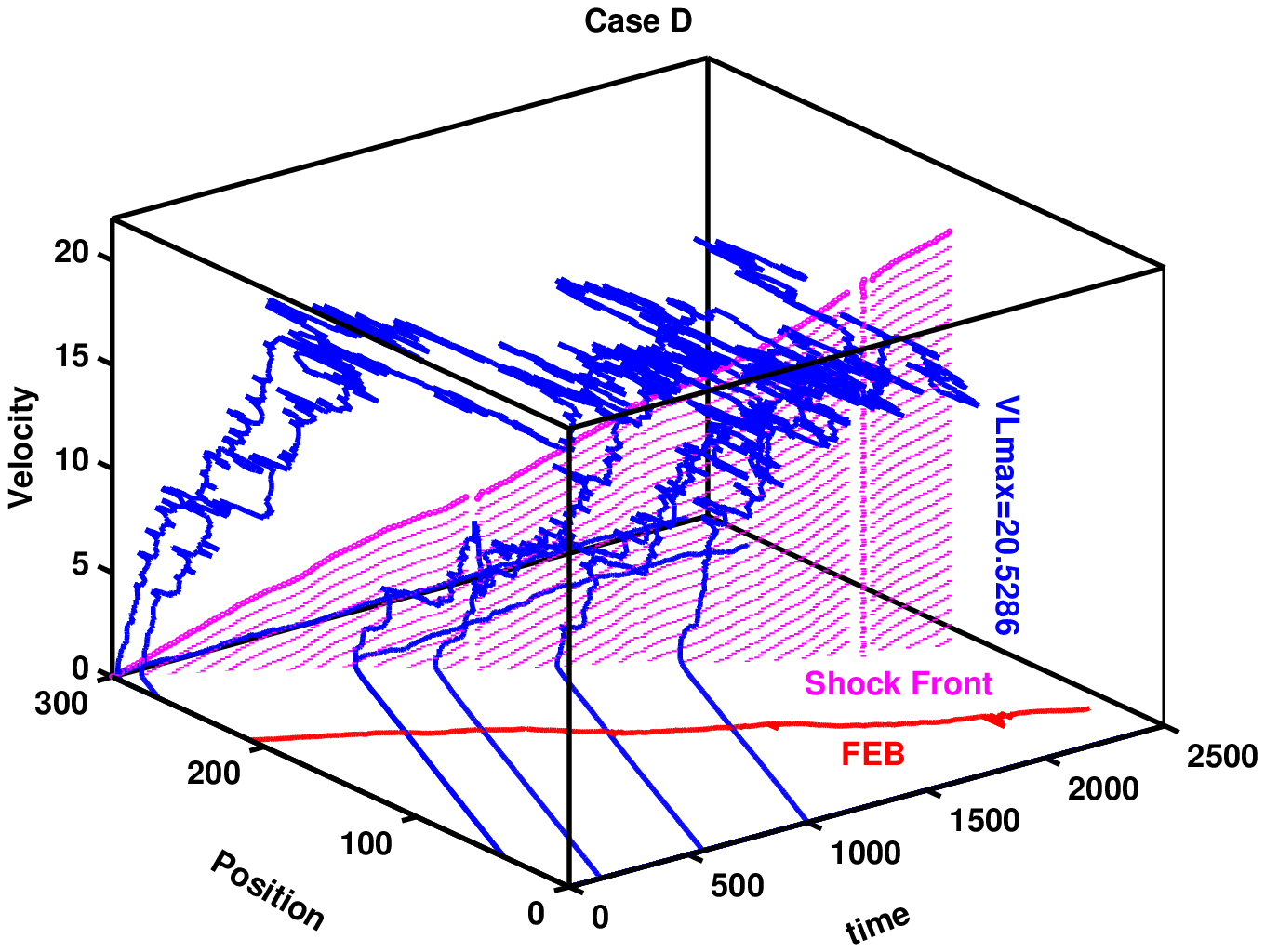}
  \caption{The individual particles with their thermal velocity in the local frame
vs their positions with respect to time in each plot. The shaded
area indicates the shock front, the solid line  in the bottom plane
denotes the position of the FEB in each case, respectively. Some
irregular curves trace the individual particle's trajectories near
the shock front with  time. The maximum energy of accelerated
particles in each case is marked with the value of the velocity,
respectively.}\label{fig:acc}
\end{figure}

We select some individual particles from the phase-space-time
database recording the all particles' information. The trajectories
of the selected particles are shown in Figure \ref{fig:acc}. One of
these trajectories  clearly shows the route of the maximum energy of
accelerated particle which undergoes the multiple collisions with
the shock front in each case. The maximum value of the energy is
apparently different in each case with increasing values of
$(VL_{max})_{A}=11.4115$, $(VL_{max})_{B}=14.2978$,
$(VL_{max})_{C}=17.2347$, and $(VL_{max})_{D}=20.5286$ from the
Cases A, B, and C to D, respectively. And those particles with a
value higher than the cutoff energy are unavailable owing to their
escaping from the FEB. The statistical data show  the number of
escaped particles in each case decreases with the number of
particles $(n_{esc})_{A}=1037$, $(n_{esc})_{B}=338$,
$(n_{esc})_{C}=182$, and $(n_{esc})_{D}=127$ from  Cases A, B, and C
to D, correspondingly. Except for the maximum energy of the particle
in each case, the other particles show that some of them obtained
finite energy accelerations  from the multiple crossings with the
shock and some of them do not have additional energy gains owing to
their lack of probability for crossing back into the precursor due
to their small diffusive length scale. The statistical data also
exhibit that the inverse energy injected from the downstream back to
upstream is characterized by a decreasing reflux rate of
$(R_{in})_{A}=38.25\%$, $(R_{in})_{B}=25.67\%$,
$(R_{in})_{C}=19.98\%$, and $(R_{in})_{D}=14.80\%$ in corresponding
Cases A, B, C, and D. With the decrease of the inverse energy from
Cases A, B, and C to D, the corresponding energy losses are also
reduced at the rate of $(R_{loss})_{A}=22.27\%$,
$(R_{loss})_{B}=17.21\%$, $(R_{loss})_{C}=13.10\%$, and
$(R_{loss})_{D}=8.86\%$ in each case, respectively. Although the
maximum energy of the accelerated particles should be identical
because of the limitation of the same size of the FEB in four cases,
the cutoff energy values are still modified by the existence of the
energy losses in the different cases applied with the different
prescribed Gaussian scattering laws.

\subsection{Energy spectrum}\label{subsec:spectrum}
\begin{figure}[t]\center
    \includegraphics[width=2.5in]{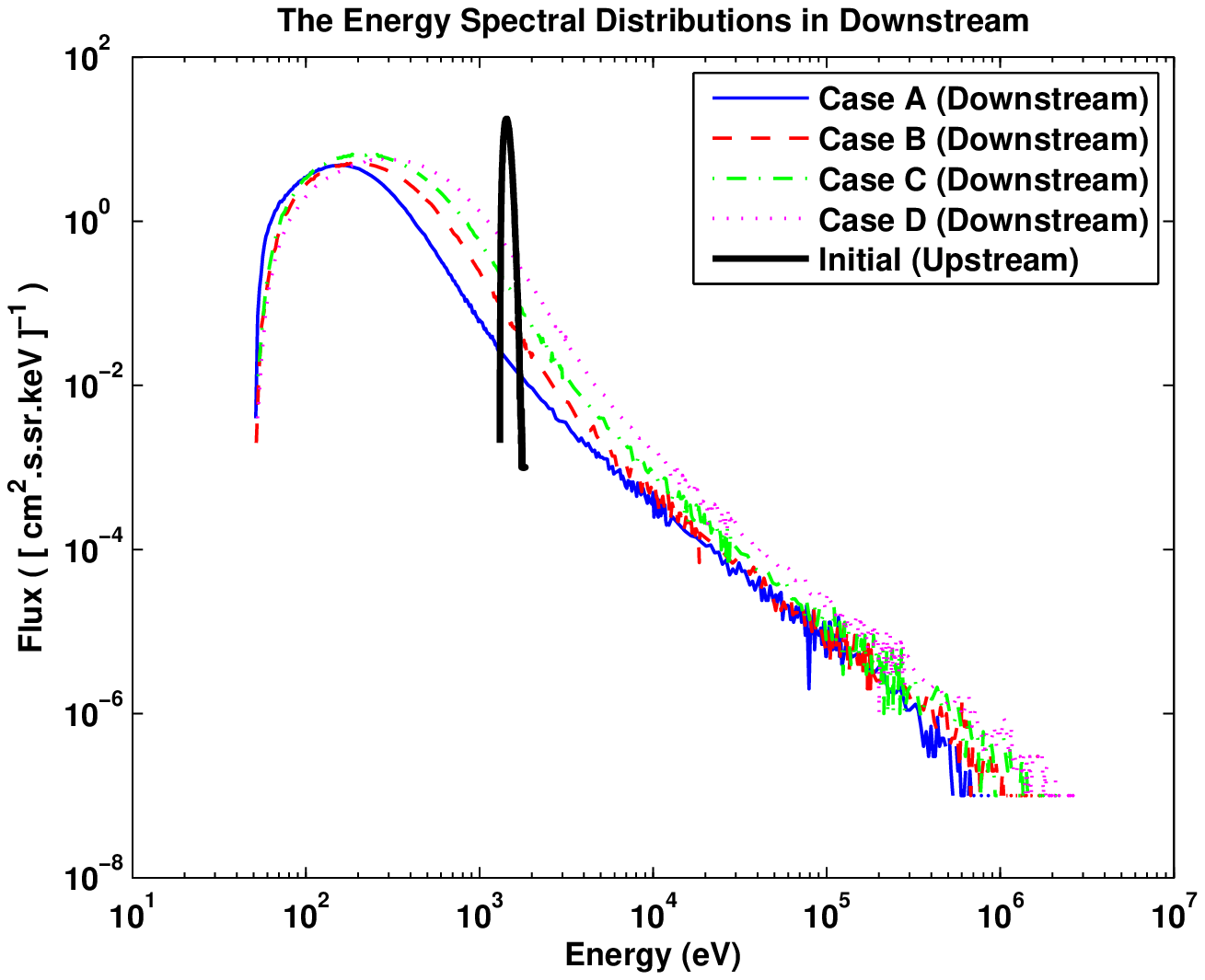}
    \includegraphics[width=2.5in]{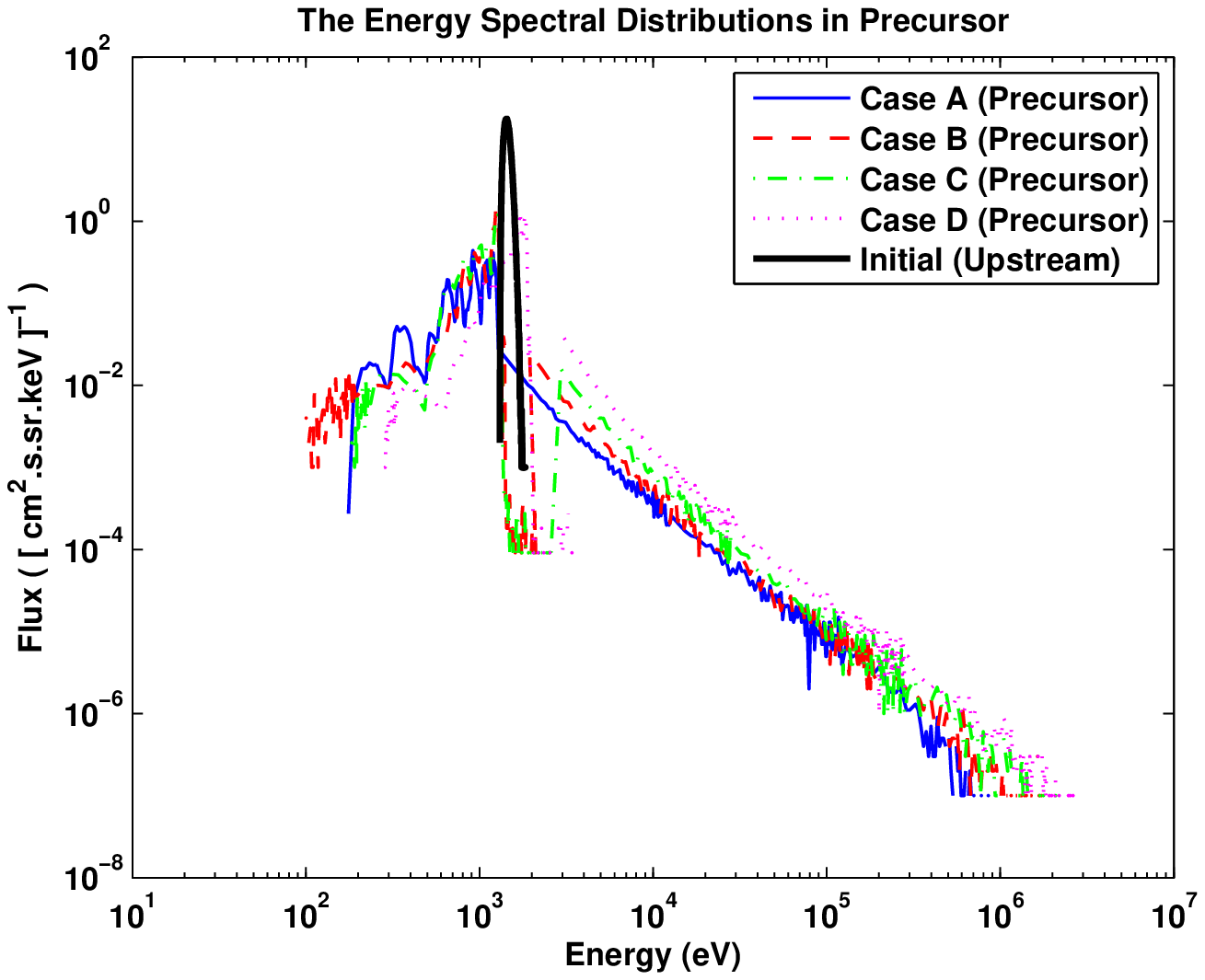}
\caption{The two plots present the final energy spectrums on the
downstream and the precursor region, respectively. The thick solid
line with a narrow peak at $E = $1.3105keV in each plot represents
the same initial Maxwell energy distributions in each case. The
solid, dashed, dash-dotted and dotted extended curves with the
``power-law" tail present the energy spectral distributions
corresponding to Cases A, B, C and D, respectively. All these energy
spectrum distributions are plotted in the same shock frame.
}\label{fig:spec}
\end{figure}

The energy spectrums with the ``power-law" tail are calculated in
the shock frame from the downstream region and the precursor region
at the end of simulation, respectively. The same initial Maxwellian
distribution in each case is shown in each plot in Figure
\ref{fig:spec}. As shown in Figure \ref{fig:spec}, the calculated
energy spectrums indicate that the four extended curves in the
downstream region with an increasing value of the central energy
peak $(E_{peak})_{A}=0.1650keV$, $(E_{peak})_{B}=0.1723keV$,
$(E_{peak})_{C}=0.1986keV$, and $(E_{peak})_{D}=0.2870keV$
characterize the Maxwellian distributions in the "heated-downstream"
from Cases A ,B, and C to D, respectively. The value of the total
energy spectral index $(\Gamma_{tot})_{A}=0.7094$,
$(\Gamma_{tot})_{B}=0.7802$, $(\Gamma_{tot})_{C}=0.8208$, and
$(\Gamma_{tot})_{D}=0.8667$ in each case indicates the Maxwellian
distribution in the "heated-downstream" with a decreasing deviation
to the ``power-law" distribution from  Cases A, B, and C to D,
correspondingly. But the value of the subshock's energy spectral
index $(\Gamma_{sub})_{A}=1.8668$, $(\Gamma_{sub})_{B}=1.2413$,
$(\Gamma_{sub})_{C}=1.1819$, and $(\Gamma_{sub})_{D}=1.0094$ present
the energy spectrum distribution with a ``power-law" tail in each
case implying there is an increasing rigidity of the spectrum from
the Cases A, B, and C to D, respectively. The cutoff energy at the
``power-law" tail in the energy spectrum is given with an increasing
value of $(E_{max})_{A} $=1.23 MeV, $(E_{max})_{B}$=1.93 MeV,
$(E_{max})_{C}$=2.80 MeV and $(E_{max})_{D}$=4.01 MeV from the Cases
A, B, C and D, respectively. In the precursor region, the final
energy spectrum  is divided into two very different parts in each
case. The part in the range from the low energy to the central peak
shows an irregular fluctuation in each case. The irregular
fluctuation indicates that the cold upstream fluid slows down and
becomes the ``thermal fluid" by the  nonlinear ``back reaction"
processes. And the other part in the range beyond the central peak
energy shows a smooth ``power-law" tail in each case.

\begin{figure}[t]\center
    \includegraphics[width=2.5in]{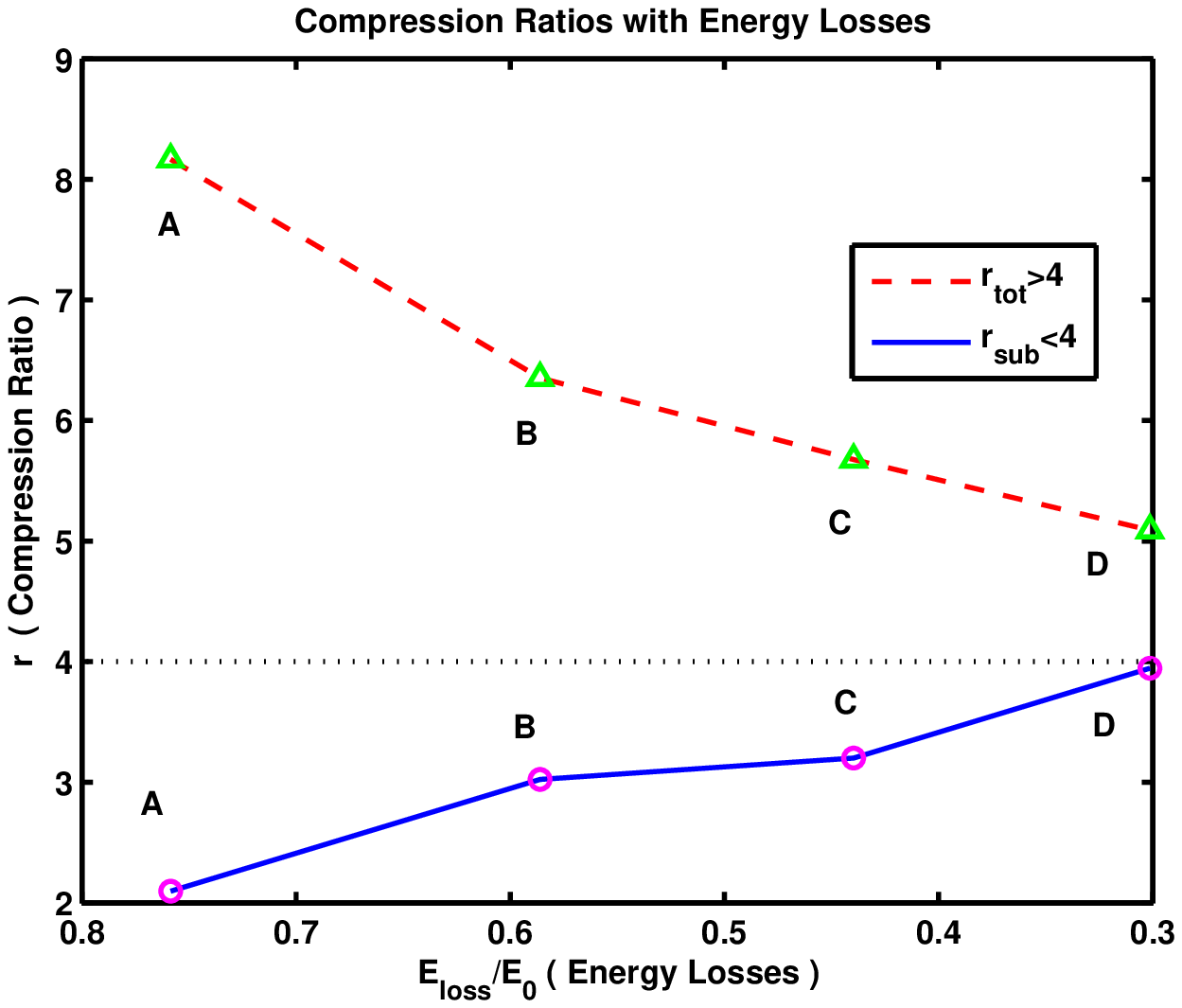}
    \includegraphics[width=2.5in]{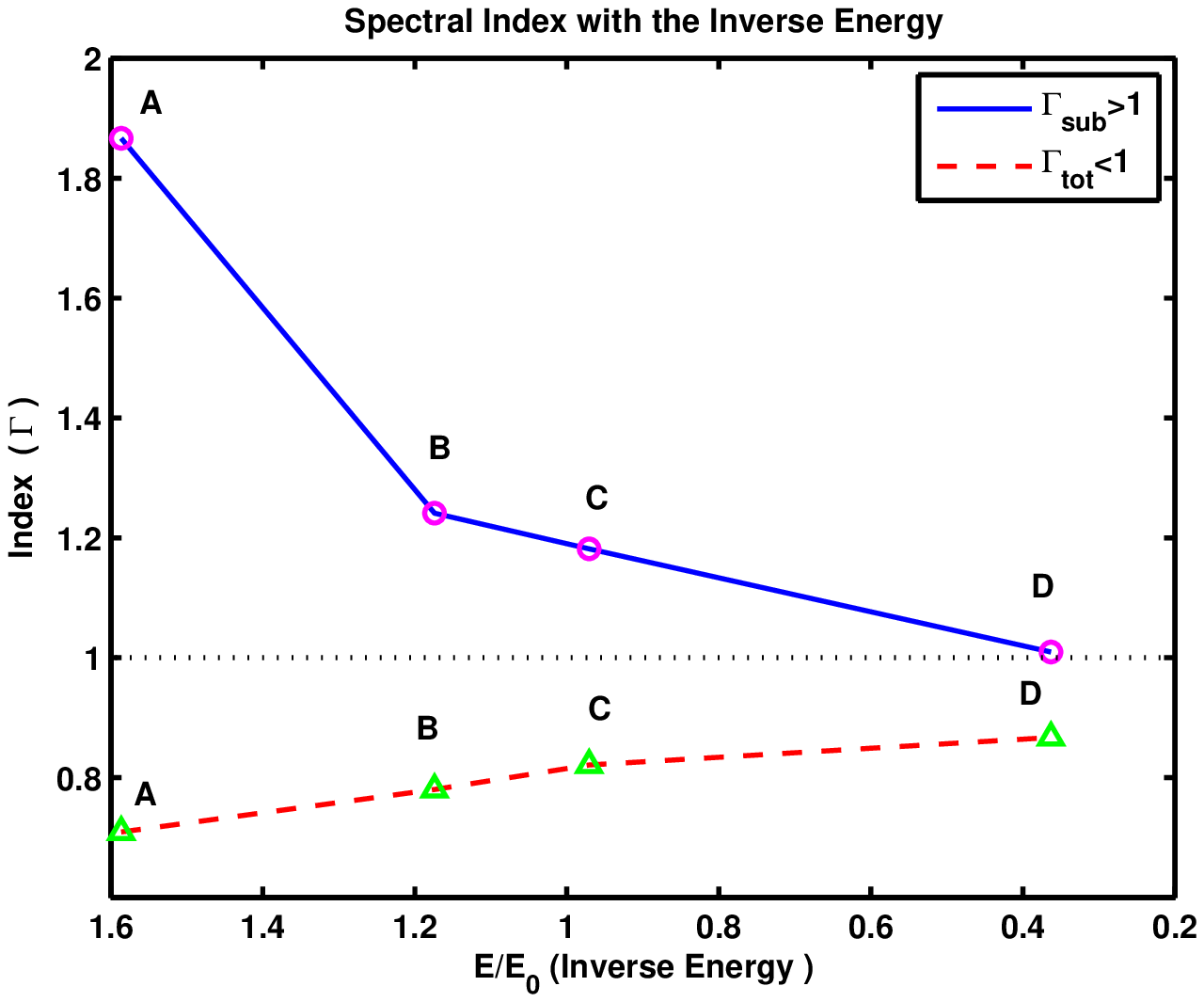}
\caption{Two plots show the correlation of the compression ratio vs
the energy losses and the correlation of the energy spectral index
vs the inverse energy, respectively. The triangles represent the
total compression ratios and the total energy spectral index of the
all cases in each plot. The circles indicate the subshock's
compression ratios and the subshock's  energy spectral index of all
cases in each plot. }\label{fig:ratio-index}
\end{figure}

As shown in Figure \ref{fig:ratio-index}, the two kinds of shock
compression ratios are both apparently dependent on the energy
losses with  respect to these four presented simulations. As viewed
from Cases A, B, and C to D, the total compression ratio is a
decreasing function of the energy losses and each value is larger
than the standard value four, the subshock's compression ratio is an
increasing function of the energy losses and  each value is lower
than four. However, both  the total compression ratio and the
subshock's compression ratio approximate  the standard value of four
as the energy loss decreases. According to the DSA theory, if the
energy loss is limited to be the minimum, the simulation models
based on the computer will more closely fit the realistic physical
situation. Additionally, the energy spectral index is also fairly
dependent on the inverse energy from the thermalized downstream
region into the precursor region.

\section{Summary and conclusions}\label{sec-summary}
In summary, we performed the dynamical Monte Carlo simulations using
the Gaussian scattering angular distributions based on the Matlab
platform by monitoring the particle's mass, momentum and energy at
any instant in time. The specific mass, momentum and energy loss
functions with respect to time are presented. A series of analyses
of the particle losses are obtained in the four cases. We
successfully examine the relationship between the shock compression
ratio and the energy losses, as well as verify the consistency of
the energy spectral index with the inverse energy injected from the
downstream to precursor region in the simulation cases which are
applied with the prescribed Gaussian scattering angular
distributions.

In conclusion, the relationship of the shock compression ratio with
the energy losses via FEB verify that the energy spectral index is
determined by the inverse energy function with time. In fact, these
energy losses simultaneously depend on the assumption of the
prescribed scattering law. As expected, the maximum energy of
accelerated particles is limited by the size of the FEB according to
the maximum mean free path in each case. However, there is still a
fairly large difference between the maximum energy of the particle
from the different cases with the same size of the FEB. We find that
the total energy spectral index increases as the standard deviation
value of the scattering angular distribution increases, but the
subshock's energy spectral index decreases as the standard deviation
value of the scattering angular distribution increases. In these
multiple scattering angular distribution simulations, the prescribed
scattering law dominates the energy losses and the inverse energy.
Consequently, the case of applying a prescribed law which leads to
the minimum energy losses will produce a harder subshock's energy
spectrum than those in the cases with larger energy losses. These
relationships will drive us to find a newly prescribed scattering
law which leads to  the minimum energy losses, making the shock
compression ratio more closely approximate the standard value of
four for a nonrelativistic shock with high Mach number in
astrophysics.

\acknowledgements{The authors would like to thank Doctors G. Li,
Hongbo Hu, Siming Liu, Xueshang Feng, and Gang Qin for many useful
and interesting discussions concerning this work. In addition, we
also appreciate Profs. Qijun Fu and Shujuan Wang, as well as other
members of the solar radio group at NAOC.}

%\bibliographystyle{apj}
%\bibliography{}

%%%%%%%%%%%%%%%%%%%%%%%%%%%%%%%%%%%%%%%%
\end{document}